\documentclass[fleqn,usenatbib]{mnras}

\usepackage{newtxtext,newtxmath}

\usepackage[T1]{fontenc}

\DeclareRobustCommand{\VAN}[3]{#2}
\let\VANthebibliography\thebibliography
\def\thebibliography{\DeclareRobustCommand{\VAN}[3]{##3}\VANthebibliography}

\usepackage{graphicx}	
\usepackage{amsmath}	

\makeatletter
\newlength{\abovecaptionskip}%
\makeatother
\usepackage{threeparttable} 
\usepackage{enumitem} 
\usepackage[dvipsnames]{xcolor}
\usepackage{soul}      
\usepackage[normalem]{ulem} 
\setstcolor{red}
\setul{}{1mm}	       

\def\hi{\relax \ifmmode {\mbox H\,{\scshape i}}\else H\,{\scshape i}\fi}
\def\hii{\relax \ifmmode {\mbox H\,{\scshape ii}}\else H\,{\scshape ii}\fi}
\def\hei{\relax \ifmmode {\mbox He\,{\scshape i}}\else He\,{\scshape i}\fi}
\def\heii{\relax \ifmmode {\mbox He\,{\scshape ii}}\else He\,{\scshape ii}\fi}
\def\nii{\relax \ifmmode {\mbox N\,{\scshape ii}}\else N\,{\scshape ii}\fi}
\def\neiii{\relax \ifmmode {\mbox Ne\,{\scshape iii}}\else N\,{\scshape ii}\fi}
\def\ni{\relax \ifmmode {\mbox N\,{\scshape i}}\else N\,{\scshape i}\fi}
\def\oi{\relax \ifmmode {\mbox O\,{\scshape i}}\else O\,{\scshape i}\fi}
\def\oii{\relax \ifmmode {\mbox O\,{\scshape ii}}\else O\,{\scshape ii}\fi}
\def\oiii{\relax \ifmmode {\mbox O\,{\scshape iii}}\else O\,{\scshape iii}\fi}
\def\sii{\relax \ifmmode {\mbox S\,{\scshape ii}}\else S\,{\scshape ii}\fi}
\def\siii{\relax \ifmmode {\mbox S\,{\scshape iii}}\else S\,{\scshape iii}\fi}
\def\ariii{\relax \ifmmode {\mbox Ar\,{\scshape iii}}\else Ar\,{\scshape iii}\fi}
\def\ariv{\relax \ifmmode {\mbox Ar\,{\scshape iv}}\else Ar\,{\scshape iv}\fi}
\def\neiii{\relax \ifmmode {\mbox Ne\,{\scshape iii}}\else Ne\,{\scshape iii}\fi}
\def\cliii{\relax \ifmmode {\mbox Cl\,{\scshape iii}}\else Cl\,{\scshape iii}\fi}
\def\feiii{\relax \ifmmode {\mbox Fe\,{\scshape iii}}\else Fe\,{\scshape iii}\fi}
\def\feii{\relax \ifmmode {\mbox Fe\,{\scshape ii}}\else Fe\,{\scshape ii}\fi}
\def\niqii{\relax \ifmmode {\mbox Ni\,{\scshape ii}}\else Ni\,{\scshape ii}\fi}
\def\cii{\relax \ifmmode {\mbox C\,{\scshape ii}}\else C\,{\scshape ii}\fi}
\def\mgi{\relax \ifmmode {\mbox Mg\,{\scshape i}}\else Mg\,{\scshape i}\fi}
\def\silii{\relax \ifmmode {\mbox Si\,{\scshape ii}}\else Si\,{\scshape ii}\fi}
\def\ha{\relax \ifmmode {\mbox H}\alpha\else H$\alpha$\fi}
\def\hb{\relax \ifmmode {\mbox H}\beta\else H$\beta$\fi}
\def\hd{\relax \ifmmode {\mbox H}\delta\else H$\delta$\fi}
\def\hg{\relax \ifmmode {\mbox H}\gamma\else H$\gamma$\fi}
\def\me{$^{-1}$}
\def\arcsec{\hbox{$^{\prime\prime}$}}
\def\arcmin{\hbox{$^{\prime}$}}
\def\deg{\hbox{$^{\circ}$}}
\def\ne{\relax \ifmmode n_e\else \hbox{$n_e$}\fi}
\def\nerms{\relax \ifmmode {\langle}n_e{\rangle}_{\mathrm{rms}}\else \hbox{${\langle}n_e{\rangle}_{\mathrm{rms}}$}\fi}
\def\sb{\relax \ifmmode \Sigma_{H\alpha} \else \hbox{$\Sigma_{H\alpha}$}\fi}
\def\SigmaSFR{\relax \ifmmode \Sigma_{\mbox{\tiny SFR}} \else \hbox{$\Sigma_{\mbox{\tiny SFR}}$}\fi}  
\defcitealias{HH09}{HH09}

\title[Electron densities and filling factors]{Electron densities and filling factors of extragalactic \hii\ regions: NGC~2403 and NGC~628}

\author[A. Zurita et al.]{
Almudena Zurita,$^{1,2}$\thanks{E-mail: azurita@ugr.es}
Fabio Bresolin,$^{3}$
Estrella Florido,$^{1,2}$
Simon Verley,$^{1,2}$
Mónica Relaño$^{1,2}$ and
\newauthor John E. Beckman$^{4,5}$
\\
$^{1}$Departamento de F\'\i sica Teórica y del Cosmos, Campus de Fuentenueva, Edificio Mecenas, Universidad de Granada, E18071--Granada, Spain\\
$^{2}$Instituto Carlos I de Física Teórica y Computacional, Facultad de Ciencias, E18071--Granada, Spain\\
$^{3}$Institute for Astronomy, 2680 Woodlawn Drive, Honolulu, HI 96822, USA\\
$^{4}$Instituto de Astrofísica de Canarias, E38205--La Laguna, Tenerife, Spain\\
$^{5}$Departamento de Astrofísica, Universidad de La Laguna, Tenerife, Spain\\
}

\date{Accepted XXX. Received YYY; in original form ZZZ}

\pubyear{\the\year{}}

\begin{document}
\label{firstpage}
\pagerange{\pageref{firstpage}--\pageref{lastpage}}
\maketitle

\begin{abstract}
Measurements of the electron density of populations of extragalactic \hii\ regions in nearby galaxies remain limited, despite the relevance of this quantity for characterizing the porosity of the interstellar medium and the escape of the ionizing radiation. We initiated a project aimed at analyzing the root-mean-square electron density \nerms, the in-situ density ($n_e$) and the volume filling factor ($\phi$) of extragalactic \hii\ regions, investigating the dependence of these attributes on nebular and host galaxy properties. We present an image-segmentation methodology for constructing homogeneous \hii\ region catalogues, and apply it to two pilot galaxies: NGC~2403 and NGC~628. We derive \nerms\ from their \ha\ luminosities and equivalent radii ($R_{\rm eq}$), and obtain $n_e$ and $\phi$ for spectroscopic subsamples. While $n_e$ is below 300~cm$^{-3}$, \nerms\ is typically one to two orders of magnitude lower, implying that $\phi$ is in the range $\sim10^{-4}$ to $10^{-1}$.  The two galaxies exhibit a similar size–density relation, $\nerms \propto R_{\rm eq}^{-0.3}$, which breaks for $R_{\rm eq}\gtrsim50$~pc, show at most a weak dependence of \nerms\ on galactocentric radius for NGC~2403, and no clear dependence of $n_e$ or $\phi$ on these parameters.
Combining these results with published data, \nerms\ presents  tentative scaling relations with the median \hii\ region size, the fraction of large regions in the parent galaxy, and the star formation rate surface density. These trends, if confirmed, would provide new constraints for massive cluster formation models and important clues for interpreting dependencies observed at high redshift, underscoring the necessity of consistently extending this analysis to larger samples.
\end{abstract}


\section{Introduction}
\label{sec:intro}
Cosmic reionization, driven by the first generations of luminous sources, transformed the intergalactic medium (IGM) from neutral to almost fully ionized, profoundly affecting its physical state and thereby constraining early galaxy formation and evolution \citep[e.g.][]{Fan2006,Robertson2015,GnedinMadau2022}. Star-forming galaxies are  considered the dominant contributors to the ionizing radiation budget, with metal-poor, low-luminosity systems  being the most likely  sources of reionizing photons \citep[e.g.][]{Flury2022,Ramambason2022,Asada2023,Mascia2024}. However, this picture remains  inconclusive, 
in part because the physical mechanisms that allow ionizing radiation to escape from galaxies are  still not well understood \citep[e.g.][]{Finkelstein2019,Naidu2020}. Nevertheless, it is clear from theoretical modelling that a key factor regulating this escape lies on small spatial scales, i.e.~in the inhomogeneous structure or porosity of the interstellar medium (ISM), which can be produced by turbulence, stellar feedback, or the presence of dense clumps, all of which may create low-density channels facilitating the leakage of ionizing Lyman continuum (LyC) photons \citep{ClarkeOey2002,Paardekooper2011,Gronke2017,Kakiichi2021,Flury2025}. 

Determining the escape fraction of LyC photons, and its dependence on galaxy properties, is crucial in order to make progress in the field. These estimates are only possible in low-redshift galaxies, which ideally should be analogues of high-redshift systems, although such a requirement can be reasonably relaxed, as the key question is to identify the conditions that allow ionizing photons to escape from galaxies \citep[e.g.][for a review]{Jaskot2025}. Before escaping from a galaxy, LyC photons must first leak out of the \hii\ regions hosting the ionizing OB associations. In nearby galaxies, there is evidence for such leakage, which, through processes involving porosity likely similar to those present at high redshift \citep{Polles2019}, allows LyC photons to escape into the ISM to ionize the diffuse ionized gas (DIG) and eventually escape into the IGM  \citep[e.g.][]{Martin1997, Zurita2000, Zurita2002, Oey2007, Belfiore2022, Watkins2024}.

Porosity in \hii\ regions is often parametrized in terms of the volume filling factor, $\phi$, implicitly assuming density fluctuations through  a simplified two-component model. In this idealization, \hii\ regions consists of multiple high density ionized clouds or clumps  that dominate the integrated line emission from the region, but do not necessarily cover all lines of sight from the ionizing source, allowing part of the ionizing radiation to escape. This concept was introduced as early as 1959, when \citet{Osterbrock1959} analysed the Orion nebula and interpreted the observed discrepancy between two independent measurements of the electron density: one obtained from emission line ratios (specifically [\oii]$\lambda$3729/[\oii]$\lambda$3726), \ne, and the much lower root mean square (rms) electron density, \nerms, also termed luminosity-weighted electron density \citep[e.g.][]{GB10}, obtained from the H$\beta$ or radio continuum surface brightness. This discrepancy was attributed to extreme density fluctuations within the region.  Subsequent observational studies have confirmed such an effect in Galactic and extragalactic \hii\ regions \citep[][]{Simpson1973,Kassim1989,HH09,Cedres2013}, substantiating the idea of a highly inhomogeneous or clumpy ISM. 
Clumpiness not only establishes a direct link between porosity, LyC leakage, and the ionization of the DIG and the IGM, but also has an impact on emission-line diagnostics, including chemical abundances, and ionization parameters \citep[e.g.][]{Giammanco2004,Giammanco2005}.

Despite its importance in studies of the ISM,  the electron density in extragalactic \hii\ regions has not received much attention so far. Although electron densities from emission line ratios have been extensively obtained for extragalactic \hii\ regions \citep[e.g.][and many more]{Zaritsky1994,bresolin2005}, measurements of \nerms\  have been obtained for much smaller samples \citep[][]{Kennicutt1984, Rozas1996, n7479_hii,Rozas00_3359,GB10, Boselli2025},
and, to our knowledge, only in a handful of cases both densities, and therefore the filling factor\footnote{$\phi$ is defined as the volume of the region occupied by the high density line-emitting clumps, relative to the total volume of the ionized nebula. Its  relation to \ne\ and \nerms\ is given by: $\phi = (\nerms/\ne)^2$. }, $\phi$, have been properly estimated \citep{Martin1997, HH09, Cedres2013}.
Determining electron densities in distant galaxies (even into the epoch of reionization) has now become more easily feasible thanks to {\em James Webb Space Telescope}  observations \citep[e.g.][]{Reddy2023,Topping2025}. The emerging results suggest that the volume filling factor of the line-emitting gas is approximately constant with redshift and that the observed increase in density with redshift may be reflecting differences in the initial conditions of the molecular clouds where stars form \citep{Davies2021,Reddy2023}. This conclusion is also supported by studies of the size–density relation of \hii\ regions, both in our Galaxy and in extragalactic systems \citep[e.g.,][hereafter HH09]{HH09}, explained as a sequence engendered by different initial gas densities, with the star-formation processes retaining an imprint from the parent molecular cloud.

The interpretation of the relations observed in high-redshift galaxies would greatly benefit from more extensive measurements of the volume filling factor and electron densities in local, spatially resolved, \hii\ regions \citep{Reddy2023,Davies2021}. Detailed studies of the ISM in the nearby universe are  critical to establish a robust connection between ISM properties and the escape of ionizing radiation, potentially helping to identify the key actors responsible for cosmic reionization. 

Our aim is to analyse the \nerms\ of the population of \hii\ regions in nearby star-forming galaxies, together with \ne\ and $\phi$ for selected sub-samples. The final goal is to investigate potential variations of these parameters with \hii\ region properties and with the properties of the host galaxies. 

A uniform methodology is essential, as the determination of \nerms\ critically depends on the adopted estimate of the ionized gas volume. Accordingly, our goal requires the derivation of \hii\ region catalogues using a consistent  approach. In this first paper we present the method and illustrate it with two pilot galaxies, NGC~2403 and NGC~628. It is organized as follows. We introduce the galaxy sample in  Sect.~\ref{sec:sample}. In Sect.~\ref{sec:data} we describe the observational data for NGC~2403 and NGC~628. The methodology employed to create the catalogues of \hii\ regions is explained in Sect.~\ref{sec:catalogue}. The derivation of \ne\ and \nerms, and their trends with \hii\ region parameters is presented in Sect.~\ref{sec:density}, and discussed in Sect.~\ref{sec:discussion}. The main conclusions are summarized in  Sect.~\ref{sec:conclusions}.

\section{Galaxy sample}
\label{sec:sample}
Our sample comprises a dozen nearby spiral galaxies with absolute magnitudes in the $B$-band between $-21$ and $-17$, at distances in the range of 2–28~Mpc, with an average of $\sim12$~Mpc, and will be fully presented and discussed in a future publication.
In this first article, we focus on NGC~2403 and NGC~628. Table~\ref{tab:galaxy_sample} summarizes some essential parameters of these galaxies. The relatively small distance to NGC~2403  \citep[3.16~Mpc,][]{Jacobs2009} introduces major challenges for automatic methods of identification and cataloguing of \hii\ regions, as the comparatively high linear resolution attainable with ground-based data permits the detection of multiple substructures and variations in surface brightness. Our second target, NGC~628 (M74), is a well-studied spiral. The combination of close proximity \citep[9.84~Mpc,][]{Distances_PHANGS_2021}, almost face-on orientation, large design spiral pattern, and presence of intense star formation has made it a preferred target for many multiwavelength galaxy surveys (e.g. SINGS, NGS, PHANGS, LEGUS, PINGS, and SIGNALS among many others). It is also arguably the galaxy for which the largest number of \hii\ region catalogues has been obtained to date (e.g. \citealt{Fathi2007}, \citealt{Cedres2012}, \citealt{RN18}, \citealt{Santoro2022}, \citealt{Groves2023}, \citealt{Congiu2023}). Nevertheless, the need to adopt a homogeneous cataloguing  procedure across our galaxy sample requires the construction of a new catalogue for NGC~628. The availability of  previous studies provides an excellent opportunity to test and compare the cataloguing methodology and to assess its impact on the derived \hii\ region parameters (Appendix~\ref{sec:appenN628}).

\begin{table}
\centering
\caption{NGC~2403 and NGC~628 adopted parameters.}
\label{tab:galaxy_sample} 
\begin{tabular}{lcc}
\hline
Galaxy                 & NGC~2403                     &   NGC~628    \\
\hline
R.A.(J2000)            & 07h36m51.3s                  &  01h36m41.8s \\
Dec.(J2000)            & +65\deg36\arcmin09.7\arcsec  &  +15\deg47\arcmin01.3\arcsec \\
D [Mpc]                & 3.16$^{a}$                   &  9.84$^{b}$ \\
$R_{25}$$^{c}$ [kpc]   & 10.0                         & 15.1 \\
$r_{e}$$^{d}$ [kpc]    & $2.9\pm0.1$                  & $5.9\pm0.9$    \\          
incl$^{d}$[deg]        & $57\pm5$                     & $32\pm3$\\
PA$^{d}$ [deg]         & $123\pm4$                    & $4\pm6$\\
$M_B$$^{c}$ [mag]      &  $-19.1$                     & $-20.2$ \\
Type$^{c}$             & .SXS6..                      &  .SAS5..  \\ 
\hline
\end{tabular}
{\footnotesize
\begin{tabular}{ll}
$^{a}$ \citet{Jacobs2009} \\
$^{b}$ \citet{Distances_PHANGS_2021}\\
$^{c}$ \citet{RC3}. $M_B$ obtained from the extinction-corrected \\
       total blue magnitude and the galaxy distance.\\
$^{d}$ \citet{PaperI}, but $r_e$ rescaled according to the distances adopted here.\\
\end{tabular}}
\end{table}

\begin{figure*}
\includegraphics[width=0.8\linewidth]{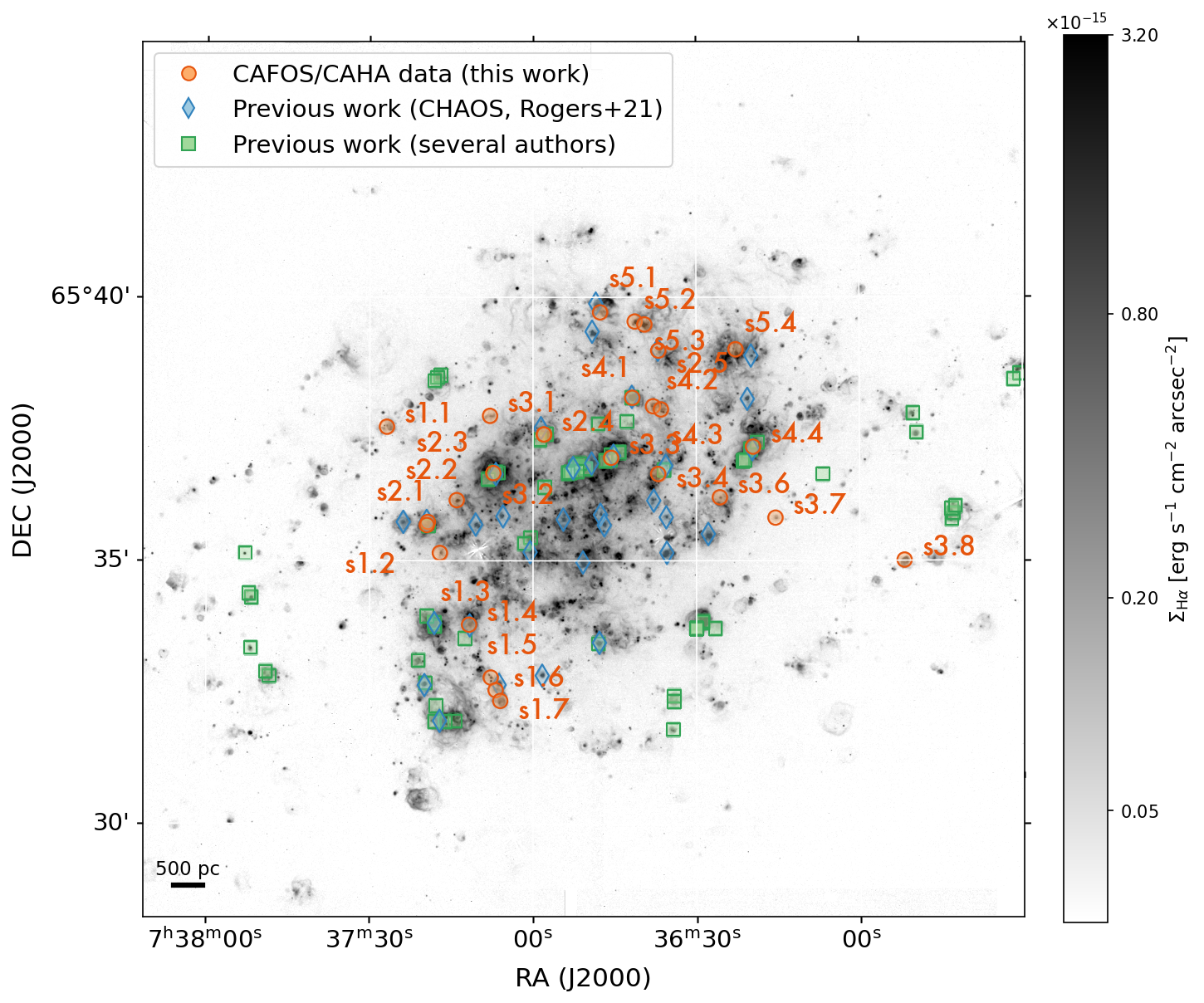}
\caption{Continuum-subtracted \ha\ image of NGC~2403 (see Sect.~\ref{sec:halpha_im}). The \hii\ regions of NGC~2403 for which optical spectroscopy has been obtained in this work are marked with orange circles, together with the corresponding identification  (see Sect.~\ref{sec:spec}). \hii\ regions for which emission-line fluxes are available from previous work are also indicated: blue diamonds for data from the CHAOS collaboration \citep{Rogers21}, and green squares for data collected in \citet{PaperI} from several authors, as indicated in Sect.~\ref{sec:compiled_data}. The colour bar shows the \ha\ surface brightness (in units of erg~s\me cm$^{-2}$ arcsec$^{-2}$).\label{fig:finderN2403}}\end{figure*}

\begin{figure*}
\includegraphics[width=0.8\linewidth]{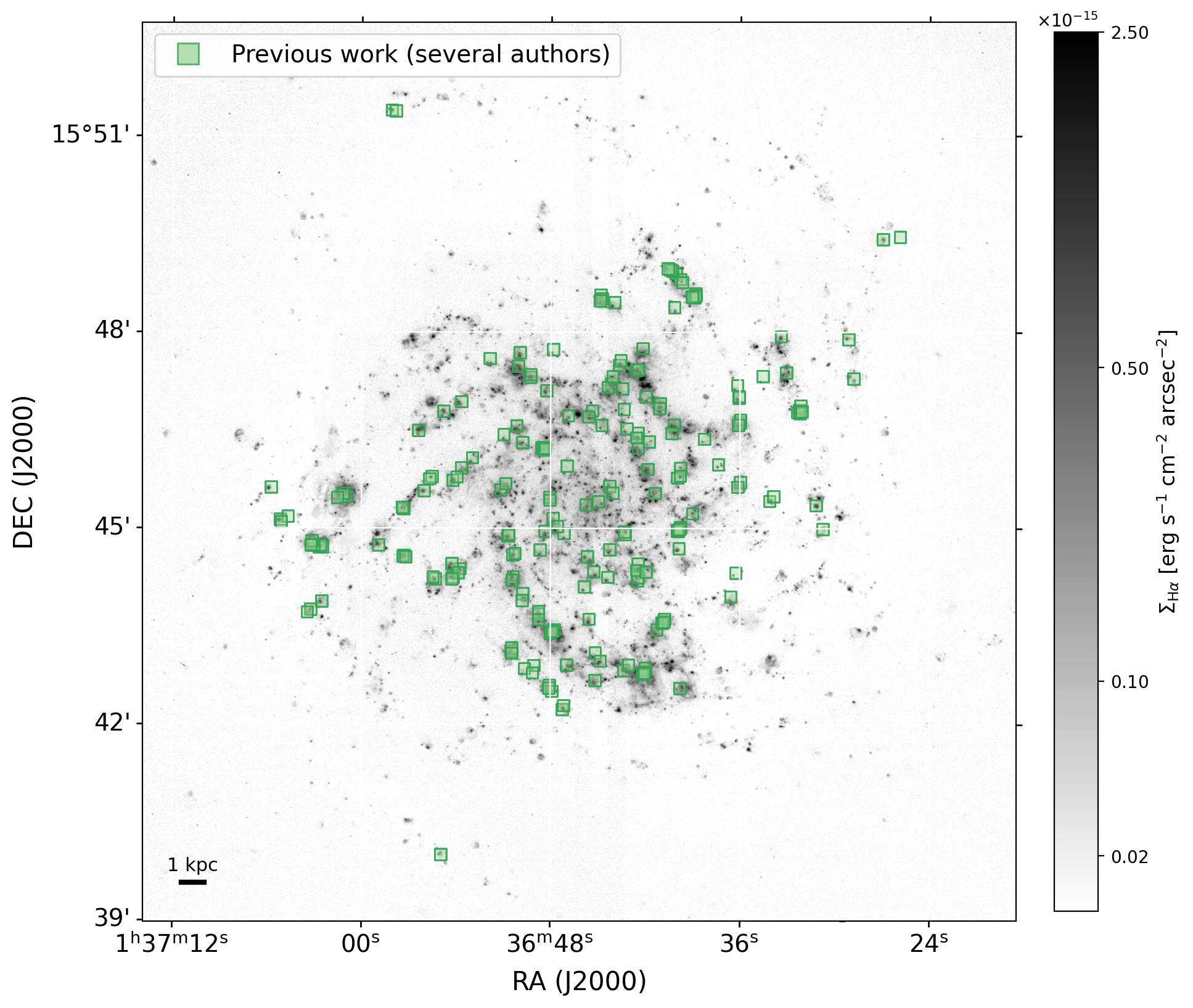}
\caption{Same as Fig.~\ref{fig:finderN2403} for NGC~628.}
\label{fig:finderN628}
\end{figure*}
\section{Observational data and data reduction}
\label{sec:data}
\subsection{Narrow-band H$\alpha$ imaging}
\label{sec:halpha_im}
The narrow-band imaging observations of NGC~2403 and NGC~628 were taken at the Roque de los Muchachos Observatory with the 4.2~m William Herschel Telescope (WHT) on the nights of 17 December and 18 September 2003, respectively. The prime focus camera, PFIP, was used for the observations. It was equipped with a mosaic of two EEV 2k$\times$4k CCDs separated by a projected $\sim$30$\arcsec$ gap  and offering a total $16\arcmin\times16\arcmin$ field of view. The camera plate scale, 0.236\arcsec\ pixel$^{-1}$, corresponds to 3.6 and 11.3~pc pixel$^{-1}$ at the distance of NGC~2403  and NGC~628 (Table~\ref{tab:galaxy_sample}), respectively. The observations consisted of exposures through a narrow-band \ha\ filter, redshifted according to the radial velocity of each galaxy (full width at half maximum (FWHM) 15.5~\AA\ and 24~\AA\ for NGC~2403 and NGC~628, respectively), and also through an off-band narrow-band filter (FWHM=15.1~\AA)\footnote{The FWHM given represent the nominal filter values, when measured in a collimated f/11 beam. The actual FWHM were respectively $\sim28$ and $21.5$~\AA\ for the on-band filters, and 21~\AA\ for the off-line filters at the f/2.8 converging beam of the PFIP/WHT instrument. The central wavelength blue shift and the reduction of the filter transmission produced by this beam were also estimated prior to the observations and considered during the data reduction and flux calibration.}. The total integration times were split into single exposures of 900~s to 1200~s for total integration times on-band (off-band) of 2400~s and 2700~s (2400~s and 4200~s) for NGC~2403 and NGC~628, respectively.
The atmospheric conditions were photometric, with seeing $\sim$0.7-0.9\arcsec\ for both galaxies.

The data reduction followed standard steps and made use of both \textsc{iraf}\footnote{\textsc{iraf} is distributed by the National Optical Astronomy Observatories, which is operated by the Association of Universities for Research in Astronomy, Inc. (AURA) under cooperative agreement with the National Science Foundation.} and  \textsc{python}/Astropy\footnote{\url{http://www.astropy.org}} \citep{astropy:2013,astropy:2018,astropy:2022}. These included bias subtraction in all frames, correction of galaxy and standard star images with appropriate flat fields and cosmic ray removal \citep[L.A.COSMIC;][]{lacos}. Next, we derived an astrometric calibration for each individual CCD image (uncertainty $\sim0.35\arcsec$). The calibration was used to correct for geometrical distortions and to create a single image from each mosaic exposure. The background sky emission was afterwards subtracted from all images, and next we aligned them, in order to combine on-band and off-band images separately, after a proper scaling to account for airmass variations between the different exposures with the same filter. Finally, the continuum off-band image was subtracted from the on-band \ha\ image.  To estimate the scaling factor of the off-band image for the continuum subtraction, we first measured fluxes of more than one hundred non-saturated foreground stars in the two images. The median ratio between the fluxes in the on-band and in the off-band images yielded an initial scaling factor value. The final scaling factor for each galaxy was determined after a careful inspection of a series of images created with scaling factors around the initial value.

Absolute flux calibration was performed via observations of the spectrophotometric standard stars Feige~34, Feige~110, BD+28-4211 and G191-B2B \citep{oke90}. 

\begin{table*}
\caption{\label{obs_log} Observing log and sky conditions.}
\centering
\begin{tabular}{lcccccccc}
\hline\hline
Slit ID  &  Slit centre         & Slit PA & N$_{\textrm{reg}}$ & Observing date  & Seeing     & Sky conditions &  \multicolumn{2}{c}{Exposure time} \\
         &  R.A. Dec. (J2000)   &   (deg) &  &                &  (\arcsec) &                &  B-100 & G-100  \\
\hline 
 N2403-s1     &  07h39m14.30s +65\deg32\arcmin08.8\arcsec  & 22.5& 7 &21/02/2020& $1.0-1.2$ & mostly clear& 2$\times$1800s & 2$\times$1800s\\ 
 N2403-s2    & 07h38m56.00s +65\deg34\arcmin34.6\arcsec  &126.8&  5&21/02/2020& $1.0-1.4$ & mostly clear& 2$\times$1800s & 2$\times$1800s\\   
 N2403-s3    & 07h38m31.10s +65\deg33\arcmin48.0\arcsec  & 70.8&  7&21/02/2020& $1.0-1.4$ & mostly clear& 2$\times$1800s & 2$\times$1800s\\    
 N2403-s4    & 07h38m23.91s +65\deg34\arcmin42.5\arcsec  & 68.1&  4 &09/02/2020 & $1.5-1.8$ & clear,cirrus& 3$\times$1800s & 3$\times$1800s\\   
 N2403-s5    & 07h38m22.86s +65\deg36\arcmin22.5\arcsec  & 74.7&  4 &10/02/2020 & $1.5-2.0$ & photometric& 2$\times$1800s & 2$\times$1800s \\  
 \hline
\end{tabular}
\end{table*}

\subsection{\texorpdfstring{Optical spectroscopy of \hii\ regions}
                           {Optical spectroscopy of H II regions}}
\label{sec:spectroscopy}
\subsubsection{CAHA/CAFOS observations of \hii\ regions in NGC~2403}
\label{sec:spec}
The spectroscopic observations were obtained with the Calar Alto Faint Object Spectrograph (CAFOS) at the 2.2~m telescope in the Calar Alto observatory in Sierra de los Filabres (Almería, Spain). CAFOS was used in long-slit spectroscopic mode, equipped with a SITe CCD with 24~$\mu$m pixels, resulting in a spatial scale of 0.53\arcsec\ pixel$^{-1}$. The observations were carried out in service mode using blocks distributed over different nights in February 2020 (9th, 10th and 21st). We used two grisms, B-100 and G-100, which provided unvignetted spectral ranges of 3200-5800~\AA\ (dispersion: 1.8~\AA\ pixel$^{-1}$) and 4900-7800~\AA\ (dispersion: 2.0~\AA\ pixel$^{-1}$), respectively. The slit width was set to 1.2\arcsec, yielding a 5.2~\AA\ FWHM spectral resolution in the two spectral ranges.

The target \hii\ regions were selected from the \ha\ image (Sect.~\ref{sec:halpha_im}), from which we defined different central positions and position angles (PA) for the 11\arcmin\ CAFOS slit. These were chosen on the basis of two main criteria: (1) to observe as many \hii\ regions as possible within a given slit position in order to extend the coverage of existing observations from the literature (Sect.~\ref{sec:compiled_data}), and (2) to have a PA matching the parallactic angle of NGC~2403 during the observations, in order to minimize light losses due to atmospheric differential refraction.  In all cases, the zenithal distance was kept below $\sim50\deg$. Five different slit positions were used, including 27 \hii\ regions,  shown in Fig.~\ref{fig:finderN2403}. Details on the slit positions, observing conditions and integration times are given in Table~\ref{obs_log}.

The data reduction was carried out with {\sc iraf} using standard routines for long-slit spectra. These included  bias and flat-field corrections, cosmic-ray rejection \citep[L.A.COSMIC;][]{lacos} and wavelength calibration of the 2D spectra using the Hg, He, and Rb lamps (with a 5th order polynomial and  rms residuals $\sim0.1$~\AA\ for the two spectral ranges). 

The 1D spectra of each \hii\ region were extracted by integrating along the spatial direction in the wavelength-calibrated 2D spectra within a defined aperture size (shown in last row of Tables~\ref{line_fluxes} and \ref{line_fluxes2}.
For each \hii\ region, the extracted 1D spectra from the B-100 and G-100 grisms were combined separately, after applying the corresponding flux calibration. The latter was performed from nightly observations
of at least two standard stars, chosen among a set which included Feige~34, BD+33-2642, BD+75-325, and G191-B2B \citep{oke90}, obtained through a 11.7 arcsec-wide slit. 

Our main aim, the derivation of electron densities, only requires the measurement of the flux of the [\sii]$\lambda\lambda$6717,~6731 doublet lines. However, as the present observations enlarge the sample of \hii\ regions observed spectroscopically in NGC~2403, we have measured the fluxes of all the emission lines detected in the CAFOS spectra. We used the {\sc IRAF} routine {\sc splot} by fitting a single (or a multiple, in the case of blending, e.g. for [\sii]$\lambda\lambda$6717,~6731) Gaussian profile. The emission lines in common between the B-100 and G-100 grisms ([\oiii]$\lambda$4959, [\oiii]$\lambda$5007, and \hei\,$\lambda$5876) were used to verify the consistency of the relative flux calibration, for which we estimate uncertainties $\lesssim7$\%.

The observed line fluxes were corrected for interstellar extinction by determining the extinction coefficient {\it c}(\hb)  from the intensities of the Balmer series hydrogen recombination lines. The \citet{howarth} parametrization of the interstellar reddening law by \citet{seaton} was adopted for $R_V=3.1$. As the measured Balmer line fluxes may be affected by absorption by the  stellar population underlying the \hii\ region emission, a correction may need to be applied. We assumed that the equivalent width of the stellar absorption (EW$_\mathrm{abs}$) is the same for all Balmer lines of a given \hii\ region, and followed the process described in \citet{zb12} to determine both {\it c}(\hb) and EW$_\mathrm{abs}$. In short, we determine the combination of {\it c}(\hb)  and EW$_\mathrm{abs}$ that produces agreement between the observed Balmer flux line ratios \ha/\hb, \hd/\hb, and \hg/\hb, and those expected  for the case B recombination given by \citet{storey95} for an electron temperature of 9000~K\footnote{The median value of the electron temperature for the low and high ionization zones in those \hii\ regions where it could be determined (see Sect.~\ref{sec:compiled_data}) is, respectively, ($9000\pm1000$)~K and ($8700\pm1600$)~K, where the error represents the standard deviation of the values.} and electron density of 100~cm$^{-3}$.

The reddening-corrected emission line fluxes of the strongest lines, normalized to \hb\,=\,100, are presented in Tables~\ref{line_fluxes} and \ref{line_fluxes2}. The errors reported account for uncertainties in the observed fluxes (calculated as in  \citealt{berg13}, considering the rms noise in the continuum around the line profile, and the uncertainty associated with the flux calibration from standard stars, estimated to be 2\%), and for the uncertainty in the derived value of {\it c}(\hb). Only lines with a signal-to-noise ratio greater than 3 are included in the tables and considered in the analysis.
\subsubsection{Compiled published spectroscopic data}
\label{sec:compiled_data}
Previous spectroscopic studies have observed \hii\ regions in NGC~2403 and NGC~628. Many of these were revisited by \citet{PaperI}, who made a compilation of optical emission line fluxes and positions of \hii\ regions in nearby galaxies from published papers, which included a total of 77 observations of \hii\ regions in NGC~2403 \citep[including data from:][]{smith75,Mccall85,Fierro86,Garnett97,vanZee98,Bresolin99,esteban09,berg13,Mao18} and 210 in NGC~628 \citep[including data from:]
[]{Mccall85,vanZee98,Ferguson98,Bresolin99,Castellanos02,Rosales-Ortega11,Gusev12,berg13,Berg15}. In addition, more recently, \citet{Rogers21}
studied \hii\ regions in NGC~2403 from spectroscopic data obtained with the Large Binocular Telescope within the CHAOS collaboration. Their sample includes 33 \hii\ regions, for which we have also collected the emission line fluxes.

The complete spectroscopic sample utilized in this work comprises 137 observations of optical emission-line fluxes of \hii\ regions in NGC~2403 and 210 in NGC~628, some of which are repeat observations of the same objects. The [\sii]$\lambda\lambda$6717, 6731 doublet, from which we derive the in-situ electron density (Sect.~\ref{sec:in-situ-density}), is included in 89 observations in NGC~2403 and 164 in NGC~628 (again with some repetitions).

\section{\texorpdfstring{\hii\ region catalogue}{H II region catalogue}}
\label{sec:catalogue}
The derivation of \nerms\ requires estimates of both the \ha\ luminosity and the volume  of \hii\ regions (see Sect.~\ref{nerms}). This implies determining the boundaries of the \hii\ regions from narrow-band images, together with assumptions on the 3D shapes of the ionized regions from their projected appearance on the plane of the sky.
However, determining such boundaries  is complicated for different reasons. For example, the spatial resolution and the signal-to-noise ratio of the observations may limit the detection of
both the faintest regions and the outermost, low surface brightness isophotes of brighter \hii\ regions, thereby affecting the extent of their projected areas on images.
In the case of deep imaging, the detection of diffuse ionized gas (DIG) adds complexity to the definition of the projected area, as there is no sharp edge around the regions.

In addition to these observational issues, measurements of \hii\ region luminosities and sizes strongly depend on the methodology and the specific identification criteria adopted by the user. These refer to general assumptions made about the expected morphology of the \hii\ regions on imaging data, which is diverse and much more  complex than that of spherically symmetric Str\"omgren spheres. As an example, for some authors, each \ha\ peak in an image can be considered as an individual \hii\ region  \citep[e.g.][]{HIIphot, RN18, bresolin25, McClain25}. Alternatively, regions can also be defined according to an \ha\ surface brightness threshold locally defined to account for the diffuse ionized gas emission, allowing for the presence of multiple \ha\ intensity peaks within the projected area of a single \hii\ region \citep[e.g.][]{Kennicutt1984, n7479_hii, Rozas00_3359}. The former approach results in catalogues containing a considerably larger number of fainter and smaller \hii\ regions than the latter.  This can lead to non-negligible differences in the properties of the \hii\ region population in a galaxy, which depend mostly on the methodology. Catalogues of \hii\ regions in the nearby spiral NGC~628 published by different authors illustrate this well (Appendix~\ref{sec:appenN628}).

In conclusion, the comparative analysis of \nerms\ for populations of \hii\ regions in different galaxies therefore requires a homogeneous cataloguing methodology. This justifies the need for new \hii\ region catalogues, even though some already exist for certain galaxies in our sample.

\begin{figure*}
    \centering
    \includegraphics[width=\linewidth]{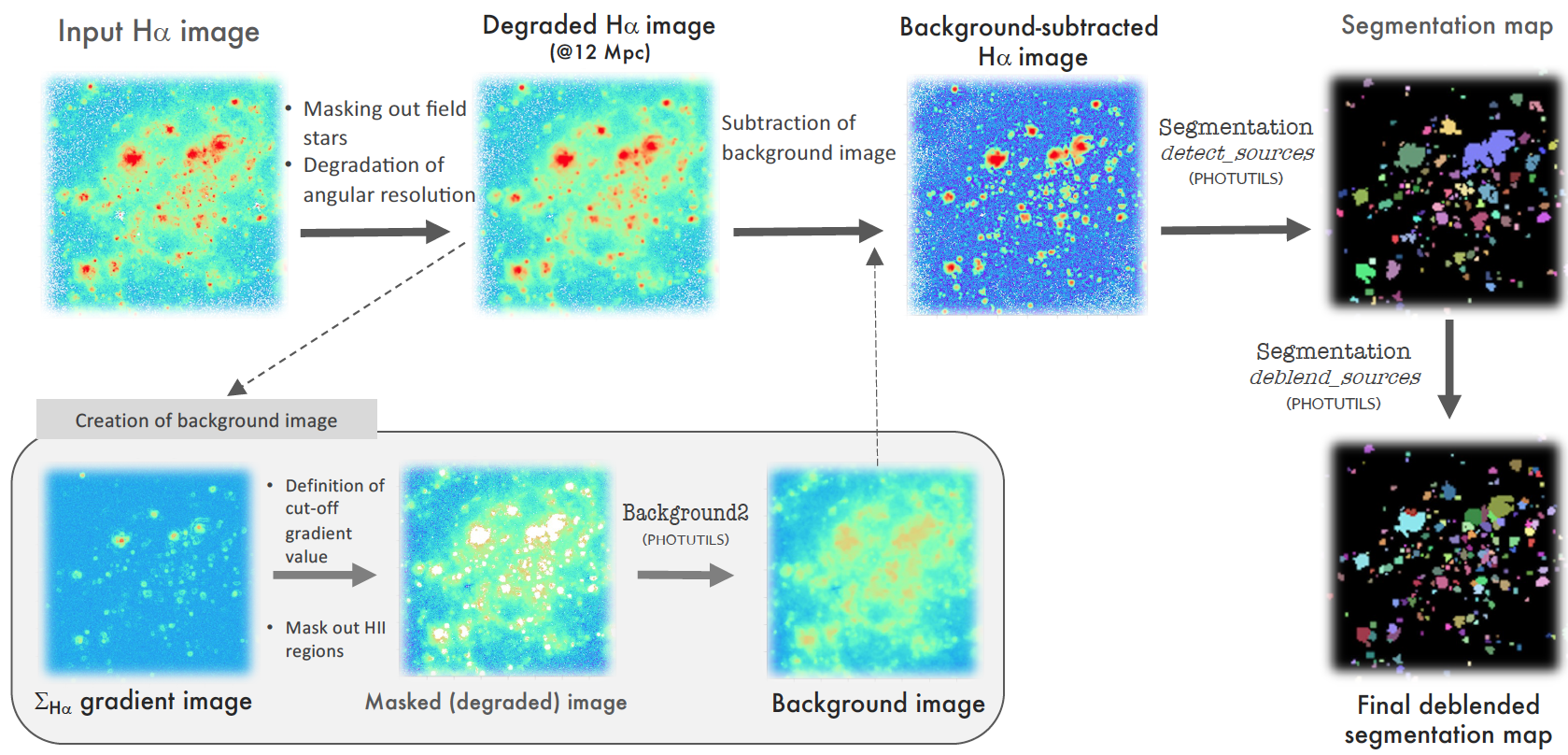}
    \caption{Flow chart diagram illustrating the method used to create a map of segments or structures in the input \ha\ image from which the \hii\ region catalogue is derived. See Sect~\ref{sec:cat_method} for details. We include the name of the algorithms in the {\sc photutils} package used to carry out the different steps.}
    \label{fig:flow_chart}
\end{figure*}

\subsection{\texorpdfstring{Methodology for \hii\ region identification: Segmentation}
                            {Methodology for H II region identification: Segmentation}}
\label{sec:cat_method}
In our work, creating the \hii\ region catalogue of a galaxy, i.e.~identifying its \hii\ regions and measuring their on-sky positions, \ha\ luminosities and projected areas, requires continuum-subtracted \ha\ imaging, and makes use of open-source astronomy packages within the Astropy project in \textsc{python}  \citep{astropy:2013,astropy:2018,astropy:2022}. For evaluation purposes, we also rely on complementary data: the emission in the continuum next to the \ha\ line (off-band narrow-band filter images, Sect.~\ref{sec:halpha_im}) and publicly available $GALEX$ images \citep{galex}, used to infer the location of the ionizing clusters.

In short,  \hii\ regions are identified as structures or {\em segments} having a defined minimum surface projected area (dependent on the spatial resolution of the data) and signal above a given intensity threshold on an artificially created image, hereinafter the {\em background-subtracted} \ha\ image.  The subtracted background approximates the diffuse \ha\ emission of the galaxy. The key parameter in the creation of the background image is the  pixel-to-pixel spatial variation of the \ha\ emission (i.e. the spatial gradient of the \ha\ surface brightness). The relevant steps are summarized in the flow chart in Fig.~\ref{fig:flow_chart} and explained in further detail below.
\subsubsection{Degradation of the angular resolution}
\label{resolution}
Our galaxy sample comprises galaxies in the distance range $\sim2$-28~Mpc, with only four galaxies lying further than the mean distance ($\sim$12~Mpc). In order to be able to better compare \hii\ region properties between galaxies, reducing the effect of differential physical resolution in the derived catalogues for our sample galaxies, while maintaining an acceptable level of spatial resolution for defining the \hii\ regions\footnote{At 12~Mpc, 0.8\arcsec\ corresponds to $\sim$45~pc. Typical extragalactic \hii\ region diameters are in the range $\sim$90–800~pc \citep[e.g.][]{Rozas00_3359, Oey2003}.}, we decided to use the mean galaxy distance as a reference. 
Therefore, the original on- and off-band images are degraded, after masking out field stars, to create images with the expected angular resolution of a spiral galaxy located at a distance of $12$~Mpc, observed with an atmospheric seeing $\sim$0.7\arcsec-1\arcsec, but using the same noise properties and pixel size of the original images.
The point spread function (PSF) was assumed to be Gaussian, with a FWHM equal to the average of non-saturated stars in the original \ha\ images (0.80\arcsec\ for NGC 2403 and 0.86\arcsec\ for NGC 628). The images were then convolved with a Gaussian kernel, reducing the angular resolution by factors 3.8 and 1.2 for NGC~2403 and NGC~628, respectively,
yielding angular resolutions of $\sim$3\arcsec\ and 1\arcsec. This corresponds to approximately the same physical resolution in the two galaxies. For this reason, the smallest ionized nebulae we consider in our work have diameters of order $\sim$40~pc. After smoothing, we added Gaussian random noise with a standard deviation equal to the background rms of the original images. From now on, when mentioning the \ha\ images, we will refer to these degraded versions.
It is worth pointing out that the minimum radii ($\sim 20$~pc) of the regions we consider in our work are comparable to the values encountered in recent surveys like PHANGS-MUSE (\citealt{Emsellem:2022}). Imaging with $HST$ allows to extend the analysis to smaller regions (\citealt{Barnes:2026}). Our study focuses on homogeneously determined properties of extragalactic giant \hii\ regions  with radii larger than $\sim 20$~pc.

\subsubsection{Creation of the background image} 
\label{background}
\hii\ regions are embedded in an extended emission component,  the DIG, that makes it difficult to define their boundaries. The difficulty arises from the fact that the surface brightness of the DIG changes across the galaxy discs \citep[e.g.][]{Ferguson1996,Zurita2000,Belfiore2022}, which implies that the isophotal level separating \hii\ regions from the DIG depends on the location of the region within the galaxy. Therefore, defining the \hii\ region boundaries requires more elaborate methods than just a simple fixed threshold in \ha\ surface brightness \citep[][]{n7479_hii, HIIphot, RN18}. 

Our approach consists in creating a map, hereafter {\em background image}, that traces the \ha\ surface brightness (hereinafter \sb)  of the DIG. This map is subtracted from the \ha\ image of the galaxy, giving an image that is intended to contain emission from \hii\ regions only, with minimal contribution from the DIG. The \hii\ region boundaries will then be determined from the background-subtracted \ha\ image, using a fixed threshold in \sb, valid across the whole galaxy face (Sect.~\ref{segmentation}). 

The background image is created from an observed property of the DIG: its low spatial gradient in \sb\ compared to \hii\ regions \citep[][]{Zurita2000}. The process of creating the background image is illustrated in the bottom-left part of Fig.~\ref{fig:flow_chart}, and consists of the following steps: (1) creation of a map of the \sb\ gradient from the degraded \ha\ image, in which the \hii\ regions stand out easily on a fairly flat background; (2) definition of the \sb\ gradient value (cut-off gradient) that best defines the boundaries of the \hii\ regions (after careful inspection of a selection of regions well distributed across the disc); (3) masking out pixels in the \ha\ image having \sb\ gradient above the cut-off gradient (i.e. mask out pixels clearly dominated by \hii\ region emission); (4) performing sigma-clipping statistics in boxes of given size in the previously masked \ha\ image to create a low-resolution background image. Masked pixels are interpolated in this low-resolution background.
We used the Source Extractor algorithm (SExtractorBackground, of the {\tt Background2D} class in the {\sc photutils} module of {\sc python}) to estimate the background in each box. However, the choice of algorithm does not have a major impact on the resulting background image, whereas the selected box size does. 

After careful tests with different box sizes, we found  satisfactory results (see Sect.~\ref{evaluation}) for sizes of $\sim$20-30~pc for the two galaxies, 
implying box sizes of 5 and 3 pixels for NGC~2403 and NGC~628, respectively.

Cut-off gradient values of $1.0\times10^{-17}$ and  $2.8\times10^{-18}$~erg~s\me~cm$^{-2}$~arcsec$^{-2}$~pc\me\ were used to mask out \hii\ regions in NGC~2403 and NGC~628, respectively, that in units of emission measure (EM) per parsec correspond, respectively, to 4.9 and 1.4~pc~cm$^{-6}$~pc\me. 

These values are in very good agreement with the {\em terminal gradients} (see Sect.~\ref{other_software}) reported by users of the {\tt HIIphot} cataloguing software for galaxies of similar distances as ours, such as 1.5~pc~cm$^{-6}$~pc\me\ for M51 (at roughly the same distance as NGC~628) by \citet{HIIphot}, or the values shown in Figure~3 of \citet{Oey2007} for a sample of spirals at different distances (mostly in the range $\sim$1.5-6.0 pc~cm$^{-6}$~pc\me). More recent {\tt HIIphot} users have adopted a fixed terminal gradient of  5.0~pc~cm$^{-6}$~pc\me\ \citep{Santoro2022,Groves2023}. 


\begin{figure*}
    \centering
    \includegraphics[width=\linewidth]{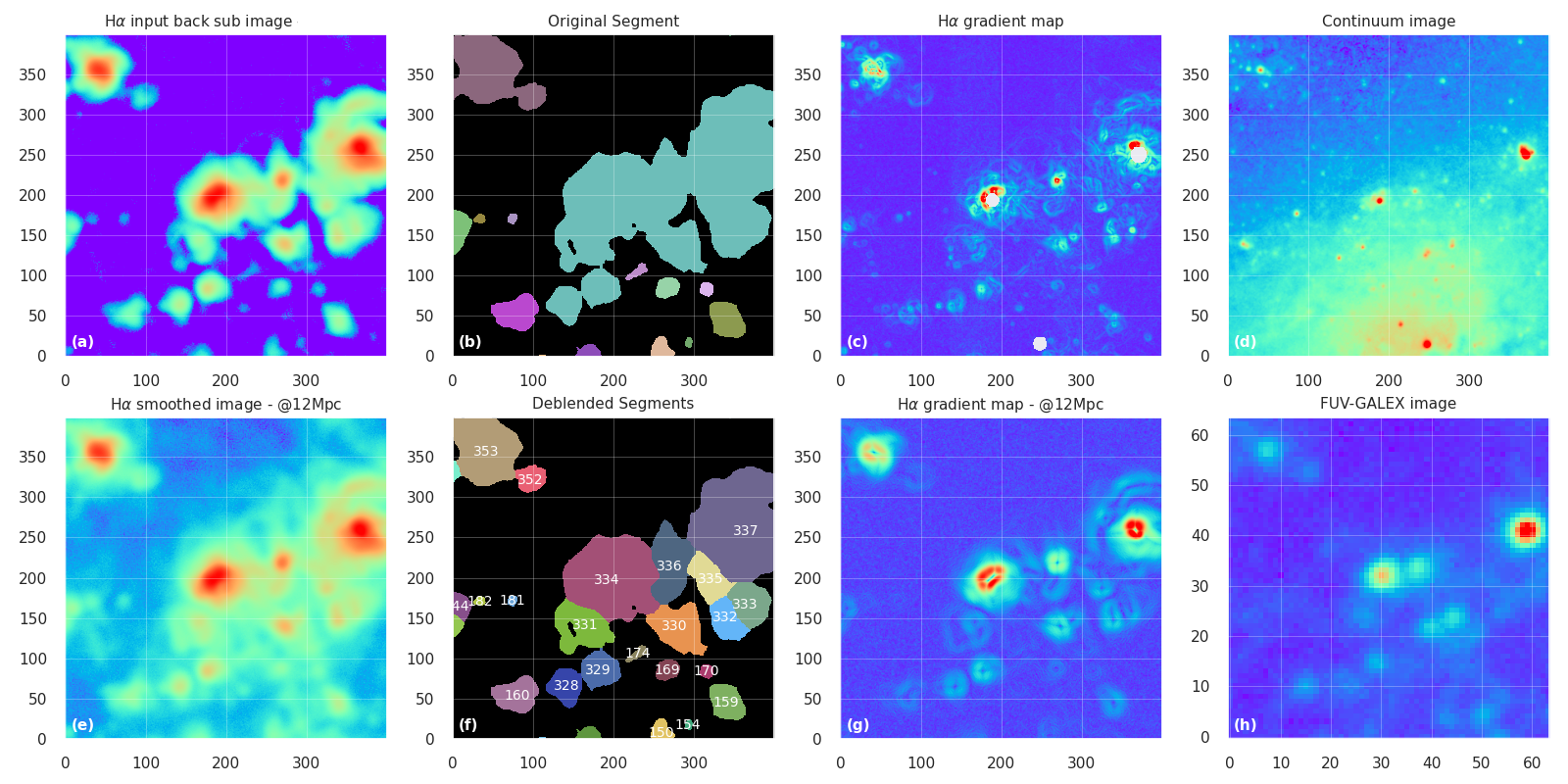}
    \caption{Portion of NGC~2403 showing  images employed in the different stages of the \hii\ region cataloguing process. {\em (a)} and {\em (e)} show the \ha\ continuum-subtracted image (with degraded resolution), with and without background subtraction, respectively (Sect.~\ref{background}).  {\em (b)} and {\em (f)} are segmentation maps before and after the deblending procedure, respectively (Sect.~\ref{segmentation}).  {\em (c)} and {\em (g)} are maps of  the \sb\ gradient from the original and degraded \ha\ image 
    (Sect.~\ref{background}), and {\em (d)} and {\em (h)} are images used to evaluate the definition of the segments and the corresponding deblending, an image of the continuum emission next to the \ha\ emission line {\em (d)}, and a FUV/$GALEX$ image of the same zone of the galaxy {\em (h)}.}
    
    \label{fig:ejemplo_evaluacion}
\end{figure*}

\subsubsection{Identification of sources: image segmentation and deblending} 
\label{segmentation}
The identification of the \hii\ regions is performed with the {\tt photutils.segmentation} subpackage\footnote{\url{https://photutils.readthedocs.io/en/2.0.2/user_guide/segmentation.html}} in {\sc python}. The \hii\ regions are defined in two steps. First, extended sources are detected in the background-subtracted \ha\ image, generated as described in Sect.~\ref{background}. These are required to have a minimum number of connected pixels with intensity  larger than three times the rms noise in this input image ({\tt detect$\_$sources} function). Such a  minimum number was set to 100 (NGC~2403) and 14 (NGC~628), equal to the number of pixels contained in a circle having the same diameter as the angular resolution of the degraded images (Sect.~\ref{resolution}).
Subsequently, a first segmentation map is created in which a label is assigned to every pixel, pixels with the same label being part of the same  source. Overlapping regions are detected as a single source in this first step. The second step consists in separating overlapping sources with a deblending procedure ({\tt deblend$\_$sources} function). This uses a combination of multi-thresholding and watershed segmentation \citep{CouprieBertrand1997}, which requires that the overlapping sources are sufficiently separated, displaying a saddle point between them. 

The parameters that control the deblending are the number of multi-thresholding levels ({\it nlevels})
and, more importantly, the {\it contrast} or minimum fraction of the total source flux that an overlapping source must have in order to be considered as a separate object. The larger the contrast value, the smaller the number of sources that will be deblended.

Intensive trials and testing allowed us to determine that an  {\it nlevels} value in the range 30-50, and a  {\it contrast} of 0.01 permit a reasonably good separation of overlapping regions at the resolution of our images (more details in Sect.~\ref{evaluation}). This contrast value implies a maximum allowed difference in $\log L_{\mathrm{H}\alpha}$ of 2~dex between overlapping regions. 
We kept the same criteria in the minimum allowed \hii\ region area as in the source detection phase also during deblending.

It is important to note that there is no single, fully objective definition of what constitutes an extragalactic \hii\ region in narrow-band images, and therefore its identification and separation cannot be uniquely determined. These depend not only on the software adopted and spatial resolution of the data \citep[e.g.][]{Barnes:2026}, but also — critically — on the user’s criteria for defining regions (as noted at the beginning of Sect.~\ref{sec:catalogue} and further discussed in Sect.~\ref{evaluation}). This issue is particularly challenging in crowded environments. Adopting clear and homogeneous criteria is therefore essential when the goal is to compare results across multiple galaxies.


\subsubsection{Generation of the catalogue} 
\label{catalogue_generation}
The \hii\ region catalogue is created from the segments map (Sect.~\ref{segmentation}) and the \ha\ galaxy image, with the {\tt SourceCatalog} class ({\sc photutils}). The catalogue contains the photometry, the location, and morphological properties (as seen in \ha) of all the sources  defined in the segmentation map.

The location is given as the coordinates of the calculated centroid within the \hii\ region segment (not necessarily coincident with the brightest point within the \hii\ region area or segment). The \hii\ region area in squared arcseconds is also included in the catalogue, together with the equivalent radius $R_{\mathrm{eq}}$ (the radius of a circle with the same area as the source segment) and other calculated morphological properties (semi-major and semi-minor axes, ellipticity and eccentricity).
The \ha\ flux of each region results from the sum of the pixel fluxes within the source segment on the \ha\ (degraded) continuum-subtracted image, and the corresponding flux uncertainties are derived from the quadratic sum of the total errors over the pixels within the segment.

Regarding the \ha\ fluxes (and luminosities) of the \hii\ regions, we have assessed the contamination due to DIG emission within the segments.
The local DIG background value (per pixel) was estimated using a rectangular annulus aperture around each \hii\ region, after masking out all pixels occupied by  catalogued \hii\ regions. The annulus width was set to ~6\arcsec\ and 2\arcsec\ for NGC~2403 and NGC~628, respectively (i.e. approximately twice the angular resolution of the degraded images), which corresponds to $\sim$80~pc. This choice is a reasonable compromise, that allows for a good sampling of the nearby DIG emission without mapping the local background at distances that may not be representative of the associated region.

The contribution of the local DIG background emission to the \ha\ luminosity can be quite high (reaching $\sim$70\%) for some of the lowest luminosity and smallest \hii\ regions, with a median contribution of $\sim$20\% for $\log L_{\mathrm{H}\alpha}\lesssim 38$~dex (or $R_{\mathrm{eq}}\lesssim 90$~pc). For larger and brighter \hii\ regions the median contribution is $\sim$10\%, with maximum values  $\sim$35\%
for a few specific regions.
The final catalogue reports $L_{\mathrm{H}\alpha}$ values that are both corrected and uncorrected for the
presence of the DIG contamination (more details in Sect. ~\ref{DIG_correction}).

 

 

The contents of the \hii\ region catalogues for NGC~2403 and NGC~628 are described in Sect.~\ref{sec:catalogues} and are available at the CDS.
\subsection{Some considerations on the cataloguing process} 
\label{considerations} 
\subsubsection{Evaluation of results} 
\label{evaluation}
The process of cataloguing \hii\ regions, like others in astrophysics, involves a certain degree of subjectivity. Publicly available software for automatic cataloguing usually gives fast and repeatable results. However, subjectivity lies in the criteria implemented in the software itself, and in how the definition of an \hii\ region is formulated within the program. In particular, it appears in the way the region boundaries are defined (which requires considerations on the DIG emission properties), and, possibly more importantly, in whether to allow for the possibility of inhomogeneities or multiple emission peaks within the \hii\ region area. 

Although we are aware that subjectivity cannot be fully removed, we made great efforts to ensure that the criteria implemented for the cataloguing, as well as our evaluation of the results, are founded on observed and intrinsic properties of the \hii\ regions and the DIG: (1) \hii\ region boundaries should be coincident with a clear change in the \sb\ spatial gradient, and (2) an \hii\ region is a volume of gas in space ionized by massive stars belonging to an OB association.

Evaluation of (1) is rather straightforward, by means of a map of the spatial gradient in \sb. A careful inspection of such a map permits us to check whether the segment's boundaries coincide with a clear change in this parameter in zones of the galaxy with different contributions from the DIG (see Fig.~\ref{fig:ejemplo_evaluacion} for an example).
The assessment of (2) is more difficult, partly because the definition of what is and looks like an OB association is  complex \citep{Wright2020}. 
However, the NUV ({\em HST}+WFPC3, $\sim$2300-3100\,\AA) emission from young stellar clusters has been found to better identify the ionizing sources of \hii\ regions compared to methods based on OB associations (\citealt{Scheuermann2023}). Proceeding in a similar fashion, we used FUV ($\sim$1350-1750\,\AA) images from {\em GALEX} (\citealt{galex}) to inform us about the presence of ionizing clusters within the \hii\ region boundaries.
The linear resolution of our degraded \ha\ and continuum images is $\sim$40-45~pc, and the FUV/{\em GALEX} angular resolution (4.2\arcsec) corresponds to $\sim$65~pc at the distance of NGC~2403 ($\sim$200~pc for NGC~628). Therefore, we have evaluated the adequacy of the deblending or possible splitting of an \ha\ emitting zone into several distinct \hii\ regions, by inspecting the location of the ionizing clusters in FUV/{\em GALEX} images and on  images of the continuum emission next to \ha, as shown in Fig.~\ref{fig:ejemplo_evaluacion}.
Our final catalogue was obtained after many tests with different parameters. As an example, when the internal substructure of the \hii\ regions coincides with the presence of multiple OB associations according to these images, the contrast parameter controlling the deblending has been lowered.
\subsubsection{Local DIG emission correction}
\label{DIG_correction}
As explained in Sect.~\ref{catalogue_generation}, we have made an estimate of the \ha\ emission (per pixel) from the DIG close to each catalogued \hii\ region. This estimate is normally scaled to the total projected area of the \hii\ region and subtracted from its integrated luminosity (in \ha\ or in any other emission line when dealing with Integral Field Spectroscopy data) in order to decontaminate the \hii\ region from DIG emission.

However, it is important to keep in mind that applying the local background correction to the \hii\ region luminosity implicitly assumes that along the line of sight, in the emitting column of each detected \hii\ region, there is significant and non-negligible DIG flux that adds up with the flux emitted by the region. Furthermore, it is assumed that the DIG contribution  is equal to the emission produced by the DIG in a column outside of the \hii\ region projected area. As the \hii\ region itself may occupy a significant fraction of the \ha\ emitting column, the use of the nearby, surrounding surface brightness of the DIG  to correct the \hii\ region luminosity may be giving a lower limit to the true \ha\ luminosity of the region, while the uncorrected value could be considered as an upper limit. A more realistic correction would require accounting for the three-dimensional distribution of the emitting gas, taking into account the \hii\ region sizes, and making additional assumptions, for example on the vertical scale height of the ionized gas component of the galactic disc, and its emissivity variation with height, which would introduce further uncertainties. For these reasons we have opted to retain the two values of $L_{\mathrm{H}\alpha}$ in the catalogue (corrected and uncorrected for the local background DIG emission) in order to keep track of possible relationships that may be affected by the correction.
\subsubsection{Limitations} 
\label{limitations}
Catalogues obtained using the methodology described above have the limitation of excluding \hii\ regions characterized by a low spatial gradient in \sb. These are usually located in the outermost parts of galaxy discs, where there is less emission from the DIG, and would be rather easily detected above a fixed emission threshold. However, these are typically very low-luminosity \hii\ regions, situated below the completeness limit. Their exclusion has a negligible impact on the statistical properties of the catalogues.
\begin{table*}
\begin{threeparttable}
\caption{Statistics of catalogued \hii\ regions in NGC~2403 and NGC~628.}
\label{tab:catalogues_stat}
\centering
\begin{tabular}{cr|ccccc|ccccc|l}
\hline\hline
Galaxy   &  N    & \multicolumn{5}{|c|}{$\log~$L$_{H\alpha}$ [erg~s\me]}   & \multicolumn{5}{c|}{$R_\mathrm{eq}$ [pc]} & Reference/Method \\
NGC~\#   &       & Min  & Max & 10\%& 50\% & 90\%                        &  Min  & Max & 10\%& 50\% & 90\% & \\
\hline
2403\tnote{a} &  370  & 36.1& 39.8& 36.3& 37.2& 38.2 & 21 & 271 & 23& 47& 95  & This work/{\tt segmentation}\\
628\tnote{a}  &  1458 & 36.0& 39.3& 36.3& 37.1& 38.1 & 24 & 314 & 32& 61& 126 & This work/{\tt segmentation}\\
\hline 
628\tnote{a,b}  & 376 & 37.2 & 39.5 & 37.3 & 37.7 & 38.6 & 36 & 213 & 38 & 56  & 112 & \citet{Fathi2007} / {\sc region} \\
628\tnote{b}  &  96 & 37.2 & 40.3 & 37.9 & 38.7 & 39.4 & 64 & 942 & 64 & 153 & 323 & \cite{Rosales-Ortega11} / Custom \\
628           & 209 & 37.5 & 40.5 & 37.8 & 38.3 & 39.1 & 38 & 339 & 41 & 69  & 155 & \citet{Cedres2012} / {\sc focas-iraf} \\     
628\tnote{b} & 4285& 31.8 & 39.5 & 36.3 & 37.0 & 37.8 & 1  & 328 & 42 & 72  & 122 & \citet{RN18}/ Custom   \\        
628           & 2137& 35.5 & 39.3 & 36.2 & 36.7 & 37.7 & - & - & - & - & - & \citet{Santoro2022} / {\tt HIIphot}\tnote{c} \\          
628& 2796\tnote{d} &  35.5 & 39.7 & 36.1 & 36.9 & 37.8 & 19 & 125 & 29 & 47 & 70 & \citet{Congiu2023} / {\sc clumpfind}\tnote{e}\\ 
628          & 2855& 35.5 &  39.8 & 36.1 & 36.7 & 37.7 & 27 & 182 & 31 & 34 & 55 & \citet{Groves2023} / {\tt HIIphot}\tnote{f} \\   
\hline
\end{tabular}
\begin{tablenotes}
\item The table includes: number of catalogued \hii\ regions (N), and the following quantities for both $\log$~L$_{\mathrm{H}\alpha}$ and equivalent radius $R_{\mathrm{eq}}$: minimum/maximum values, and 10th, 50th (median) and 90th percentiles. For comparison purposes, the values obtained from published catalogues of NGC~628 are also provided. We include the adopted cataloguing methods in the last column. \item $^{\textrm{a}}$\ha\ luminosities are not corrected for extinction. \item $^{\textrm{b}}$Luminosities and diameters are scaled to a galaxy distance of 9.84~Mpc \citep{Jacobs2009}. \item $^{\textrm{c}}$Adaptation of the {\tt HIIphot} software. \item $^{\textrm{d}}$Considers only catalogued nebulae with {\tt p\_hii}$>90$ according to these authors, i.e. those with a high probability of being \hii\ regions. \item $^{\textrm{e}}${\sc clumpfind} for initial detection, afterwards custom algorithm. \item $^{\textrm{f}}$Modified version  adapted for maps derived from IFS data \citep{Kreckel2019}. 
\end{tablenotes}
\end{threeparttable}
\end{table*}

\subsubsection{Comments in relation to other software/techniques}
\label{other_software} 
There are multiple software tools for cataloguing \hii\ regions, using algorithms of varying complexity, imposing different underlying assumptions. The last column of Table~\ref{tab:catalogues_stat} succinctly illustrates this, since at least six different methods have been employed for deriving \hii\ region catalogues in NGC~628.

{\tt HIIphot} is one of the most widely used tools. Similarly to our procedure, this IDL-based software uses \sb\ as a key parameter to determine the boundaries of extragalactic \hii\ regions. These two approaches, while somewhat different,
have in  common the underlying idea that the \sb\  of the \hii\ regions rapidly declines outwards from the central peak, until it reaches the level of the DIG, where the surface brightness profile either flattens completely (zero brightness gradient) or declines very slowly (with a gradient that is equal to our cut-off value or the terminal surface brightness slope of {\tt HIIphot}). 

Despite the similarities, the properties of the \hii\ regions derived with the two methods can be quite different. This can be seen in Appendix~\ref{sec:appenN628}, where we have made a detailed comparison of the published \hii\ region catalogues of NGC~628 listed in Table~\ref{tab:catalogues_stat}. Not only can the number of objects found and the distribution of their luminosities and sizes vary substantially between different methods, but what is most striking is the effect on 
the $\log L_{\mathrm{H}\alpha}-\log R_{\mathrm{eq}}$ relationship. For a spherically symmetric \hii\ region of radius $R$ and uniform  density $n_e$ the trend between these two quantities is expected to be approximately linear, with a slope $\sim3$ \cite[][pp.146, 166]{osterbrock1989}:
\begin{equation}
\label{LVec}
L_{\mathrm{H}\alpha} \approx h \nu_{\mathrm{H}\alpha}\, n_e^2 \, \phi \, \alpha^\text{eff}_{\mathrm{H}\alpha} \, \frac{4 \pi R^3}{3} \quad, 
\end{equation}
with $h\nu_{\mathrm{H}\alpha}$ being the energy of an \ha\ photon, $\phi$ the \hii\ region filling factor, and $\alpha^\text{eff}_{\mathrm{H}\alpha}$ the \ha\ effective recombination coefficient. However, Fig.~\ref{fig:logL-logR_all} shows that this relation varies significantly between the different catalogues, with slopes that can be quite different from 3. We also remark on the fact that three catalogues obtained with the same observational data by different members of the same collaboration \citep[MUSE-PHANGS -- ][]{Santoro2022,Groves2023,Congiu2023}, two of which using methods based on {\tt HIIphot}, display notable differences (see Figs.~\ref{fig:finders_zoom_all}, \ref{fig:logL-logR_all}), affecting the derived distribution of  \nerms\ values (Fig.~\ref{fig:logN-logR_all}).

While several additional \hii\ region cataloguing software appear in the literature \cite[e.g.][]{DellaBruna2020,pyHIIexplorer2022,Lugo-Aranda2024}, it is not our aim  to make an exhaustive comparison between them. However, it is important to stress that, as demonstrated by the comparison of catalogues for NGC~628, a uniform methodology is essential for meaningful comparisons of \hii\ region properties between galaxies. 
Furthermore, we consider it equally important to check that the physical parameters of the photoionized regions obtained from any catalogue exhibit a behaviour that is consistent with the theoretical expectations, such as the luminosity–radius relation for a spherical region mentioned above.

A recent study by \citet{McClain25} in the nearby galaxy NGC~253 explicitly demonstrates how the definition of \hii\ region radii can significantly affect the inferred $\log L_{\mathrm{H}\alpha}$ -  $\log R$ relation. In their analysis, only radii defined through a surface-brightness threshold — the approach most closely resembling the one adopted in this work — recover a clear scaling relation, although the normalization depends on the chosen threshold. Within this framework, our determination of \hii\ region sizes, while inherently subject to uncertainty, is
empirically well-motivated (gradient in \sb) and yields luminosity–radius relations, consistent with the expected behaviour of photoionized regions, with high correlation and low dispersion (see Sect.~\ref{sec:catalogues} and Fig.~\ref{L-V}).

We believe that our methodology is notable for being easy to use, as it is based on freely available python software packages, and for taking  advantage of information based on observed properties of photoionized regions that are not systematically included in catalogues -- such as the change in the \ha\ surface-brightness gradient between the DIG and the \hii\ regions, or the location of the ionising clusters. The primary motivation, however, is to provide a homogeneous and consistent approach to identifying and characterising \hii\ regions across multiple galaxies. NGC~2403 and NGC~628 are presented here as pilot examples to illustrate the method.


\subsection{NGC~2403 and NGC~628}
\label{sec:catalogues}
The \hii\ region catalogues for NGC~2403 and NGC~628 comprise 370 and 1458 regions, respectively. For each \hii\ region (or segment) the catalogues report: sky coordinates, \ha\ luminosity (both corrected and uncorrected for the contribution of the local DIG background), local DIG background emission associated with the region, projected area, the corresponding equivalent radius, semi-minor and semi-major axes, ellipticity and orientation of the 2D Gaussian function that has the same second-order moment as the segment defining the \hii\ region.
\begin{figure}
    \centering
    \includegraphics[width=0.8\linewidth]{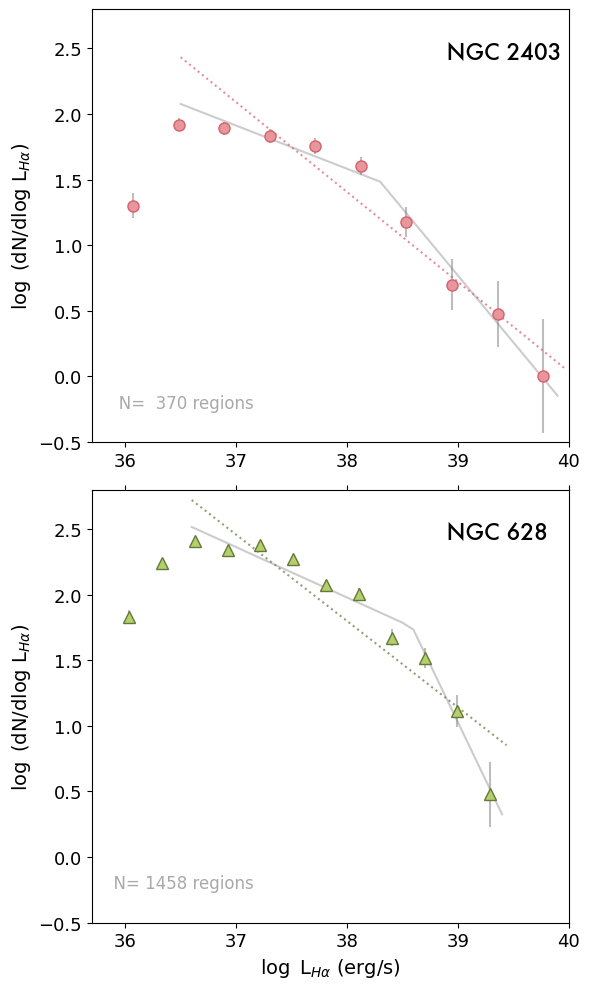}
   \caption{Differential \hii\ region luminosity functions for NGC~2403 (top) and NGC~628 (bottom). The dotted line shows a power-law fit to the data points lying above the  completeness limit, that yields similar indices for the two galaxies ($\eta=1.68\pm0.08$ for NGC~2403, $\eta=1.7\pm0.1$ for NGC~628). 
   The full lines represent fits to a double power law (see text for details).}
    \label{fig:LFs}
\end{figure}

The ranges covered by the  \hii\ region equivalent radius and the logarithm of the  DIG-corrected \ha\ luminosity are reported in Table~\ref{tab:catalogues_stat}, together with the corresponding values from previous catalogues for NGC~628, for comparison purposes (see also Appendix~\ref{sec:appenN628}). Although the minimum \ha\ luminosity is $\log~$L$_{H\alpha}$[erg~s\me]$\simeq$36.0, the completeness limit is $\log~$L$_{H\alpha}$[erg~s\me]$\simeq$36.6~dex for both galaxies\footnote{We remind the reader that the catalogues were derived from images degraded to mimic a distance of $\sim12$~Mpc (Sect.~\ref{resolution}).}. 
This can be inferred from the differential luminosity function (LF), representing the number of \hii\ regions per unit luminosity interval, shown in Fig.~\ref{fig:LFs}. The  decline in the number counts at low luminosities approximately marks the \ha\ luminosity below which the \hii\ region census becomes incomplete \citep[see][for a discussion of the factors affecting the turnover point]{Santoro2022}.

Pioneering work by \citet{KennicuttHodge80} proposed the use of \hii\ region LFs  as a diagnostic to constrain the upper end of the stellar initial mass function and the mass distribution of ionizing star clusters. This study, together with subsequent work, has empirically established that, above the completeness limit, the LF is well described by a power law of the form $dN(L_{\mathrm{H}\alpha})/dL_{\mathrm{H}\alpha} \propto L_{\mathrm{H}\alpha}^{\eta}$, with an index  $\eta=-2.0\pm0.5$, where the quoted range is not an uncertainty but rather reflects the observed galaxy-to-galaxy variations \citep[e.g.][]{KEH1989,Rand1992,Knapen1993,Rozas1996,n7479_hii,Azimlu2011,Santoro2022}.
A single linear fit to the LFs in Fig.~\ref{fig:LFs} yields $\eta\approx -1.7$, which lies well within this range, and in excellent agreement with the value $\eta= -1.71\pm0.1$ derived for NGC~628 by \citet{Santoro2022}.

The LFs in Fig.~\ref{fig:LFs} display a clear steepening for $\log L_{\mathrm{H}\alpha} \gtrsim 38$.  A double power law provides a better fit, yielding values of $\eta = -2.0 \pm 0.1$ and  $-2.8 \pm 0.3$ in the high-luminosity regime for NGC~2403 and NGC~628, respectively. The break  in the \ha\ LF occurs at $\log~L_{\mathrm{H}\alpha}$[erg~s\me]$=38.3\pm0.2$ for NGC~2403 and  $38.6\pm0.1$ for NGC~628.  The presence of this feature, typically found around  $\log~L_{\mathrm{H}\alpha}$[erg~s\me]$\simeq38.6$ \citep{Bradley2006}, 
has been attributed to several physical mechanisms, including evolutionary effects associated with a maximum number of ionizing stars per nebula \citep{OeyClarke1997,OeyClarke1998}, and a change from ionization-bounded to density-bounded regimes in the most luminous \hii\ regions \citep{Beckman2000, Bradley2006}.


We note that the \ha\ luminosities reported in the catalogues are not extinction-corrected. Using the compiled spectroscopic data for the two galaxies (Sect.~\ref{sec:compiled_data}), we estimate an average Balmer extinction of {\it c}(\hb)$=0.33\pm0.20$ and $0.35\pm0.21$ for NGC~2403 and NGC~628, respectively. The distribution of {\it c}(\hb)  values implies that about 95\% of the regions are expected to have extinction-corrected $\log$~L$_{H\alpha}$ values up to $\sim0.4$~dex larger than reported. In what follows, unless otherwise stated, we use \ha\ luminosities —and quantities derived from them— uncorrected for extinction.

The cumulative distributions of the \hii\ region diameters follow an exponential law \citep{vandenbergh}, with characteristic diameters $D_0=(83\pm6$)~pc and $(77\pm3)$~pc, for NGC~2403 and NGC~628, respectively. This will be discussed in Sect.~\ref{sec:discussion}.

\begin{figure*}
    \centering
    \includegraphics[width=0.8\linewidth]{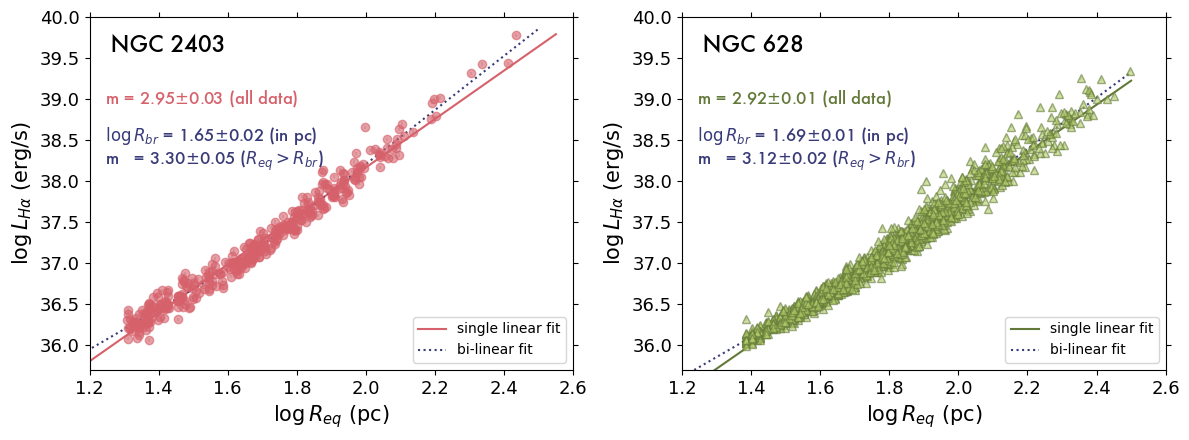}
    \caption{Logarithm of the \ha\ luminosity (in erg s\me) as a function of the logarithm of the equivalent radius (in parsecs) for the \hii\ regions of  NGC~2403 (left) and NGC~628 (right). The solid straight lines show the best linear fits to all the data points, with the measured slopes indicated in each panel. The dotted lines represent the corresponding bilinear fits, which are statistically favoured over a single linear model ($\Delta$BIC$<-10$; see Sect.~\ref{sec:catalogues}) and yield a steeper slope at large radii ($\log R_{\mathrm{eq}}\gtrsim1.7$).}
    \label{L-V}
\end{figure*}

As a final characterisation of the catalogues, Fig.~\ref{L-V} shows the \ha\ luminosity–size relation of the \hii\ regions in both galaxies. The relation is approximately linear to first order, with slopes of $2.95\pm0.03$ and $2.92\pm0.01$, for NGC~2403 and NGC~628, respectively, close to the expected value of 3 for regions with constant $\nerms$ (or, equivalently, constant $\ne^2 \phi$, see Eq.~\ref{LVec}). However, the most luminous regions show a  departure from a single power-law behaviour. A bilinear fit provides a better representation of the data, as confirmed by a model comparison using the Bayesian information criterion (BIC).  Following the criteria of \citet{KassRaftery1995} \citep[but see e.g.][for astronomy-related examples]{Liddle2007, Bresolin2020}, we adopt a threshold of  $\Delta\text{BIC} <  -10$ to justify the additional complexity of the bilinear parametrization over a single linear relation. The bilinear fit yields steeper slopes  ($\sim$3.1-3.3) at large radii, above a break radius that is similar for the two galaxies: $(44\pm2)$~pc and $(49\pm1)$~pc for NGC~2403 and NGC~628, respectively. A similar behaviour has been reported in other spirals \cite[e.g.][]{n7479_hii}.
\section{\texorpdfstring{Electron density of \hii\ regions in NGC~2403 and NGC~628}
                        {Electron density of H II regions in NGC~2403 and NGC~628}}                   
\label{sec:density}
\subsection{\texorpdfstring{In-situ electron density, \ne}{In-situ electron density, ne}}
\label{sec:in-situ-density}
We use the {\em getTemden} task in {\sc Pyneb} v1.1.18  \citep{pyneb} to calculate the electron density \ne\  for each \hii\ region for which we have spectroscopic data (Sect.~\ref{sec:spectroscopy}), using the [\sii]$\lambda\lambda6716,6731$ doublet, with the same selection of atomic data as in \citet{Rogers2022}. 
When auroral line fluxes are available, we use the electron temperature $T_e$ of the low-ionization zone, assuming a two-zone
scheme for the ionization structure of the \hii\ regions, as described in \citet{PaperI}. For the remaining cases we adopt the $T_e$ value expected from the location of the \hii\ regions within the galaxy discs, using the linear regression describing the relation between $T_e$ and the galactocentric distance\footnote{$T_e$~[K] = ($7610\pm50$) + ($450\pm180$)~$R$ [kpc] (correlation coefficient $r$=$0.77$) and $T_e$~[K] = ($6500\pm200$) + ($210\pm20$)~$R$~[kpc] ($r$=$0.82$) , for NGC~2403 and NGC~628, respectively.}. \citet{PaperI} describes the iterative process we followed, updated via the use of {\sc Pyneb}. Here we note that the absolute uncertainties in \ne\ were estimated by propagating the errors in the [\sii]$\lambda\lambda6716,6731$  line fluxes, as listed in Tables~\ref{line_fluxes} and \ref{line_fluxes2}, for the CAHA/CAFOS data. For data compiled from the literature, we adopted the flux uncertainties reported by the corresponding authors (see Sect.~\ref{sec:compiled_data}). The relative errors for both [\sii]$\lambda6716$ and [\sii]$\lambda6731$ range from 3\% to 23\%, with a mean value of $\sim 7\%$.

The electron density of the \hii\ regions in NGC~2403 was derived utilizing 89 distinct line flux measurements, and 164 in the case of NGC~628. However, several of these originate from observations of the same \hii\ regions by different authors. Based on the positional information given in the original publications, we matched all targets in the spectroscopic sample (Sect.~\ref{sec:compiled_data}) with individual segments in our \hii\ catalogue (Sect.~\ref{sec:catalogues}), thereby assigning an equivalent radius and H$\alpha$ luminosity to each in-situ electron density determination obtained from the [\sii] $\lambda6716/\lambda6731$ line ratio. This procedure obviously relies on the accuracy of the published coordinates. When multiple \ne\ estimates from different authors were available, we adopted an average value, acknowledging that the original slit pointings may probe somewhat distinct areas within the same \hii\ region. In such cases,  the uncertainty in the derived density was quantified as follows: when two measurements were available, the uncertainties were calculated  via standard error propagation; for more than two measurements, the standard deviation of the \ne\ measurements was adopted.

In total, we obtain \ne\ estimates for 36 \hii\ regions in NGC~2403 (plus 9 upper limits) and 80  \hii\ regions in NGC~628 (plus 22 upper limits).  Of these, 12 \hii\ regions in each galaxy have more than two \ne\ determinations. All regions exhibit in-situ electron densities below 300~cm$^{-3}$, with approximately 80\% in both galaxies having \ne\ below 100~cm$^{-3}$. The median value of \ne\ (excluding upper limits) is 46~cm$^{-3}$ in both galaxies, with a slightly larger dispersion in NGC~2403 (63~cm$^{-3}$) compared to NGC~628 (50~cm$^{-3}$).

The subsamples, having been selected from spectroscopic observations (mostly through long slits), naturally include some of the most luminous \hii\ regions in these galaxies. Nevertheless, they span reasonably wide ranges in $\log L_{\mathrm{H}\alpha}$ (erg\,s$^{-1}$): from 36.6 to 39.8 in NGC~2403 and from 37.2 to 39.3 in NGC~628, with median values of 38.1 and 38.4, respectively. Their radii also vary considerably, ranging from 30 to 275~pc in NGC~2403, and from 60 to 314~pc in NGC~628, with median values of 96 and 146~pc, respectively.

\subsection{\texorpdfstring{rms electron density, \nerms}{rms electron density, <ne>rms}}
\label{nerms} 
The rms or luminosity-weighted electron density, \nerms, is the density that accounts for the measured \ha\ luminosity ($L_{\mathrm{H}\alpha}$) of an \hii\ region of a given emitting volume $V$. Under  case B of the recombination of hydrogen \citep{osterbrock1989}, the total number of ionizing photons emitted by the embedded massive stars per unit time, $Q(H^0)$, balances the recombination rate to excited energy levels of the hydrogen atoms within the ionized volume ({\em Str\"omgren sphere}):
\begin{equation}
Q(H^0) =  2.23 \, L_{\mathrm{H}\alpha} = {\langle}n_e{\rangle}_{\rm rms}^2 \, \alpha_B \, h\nu_{H\alpha} \, V
\end{equation}
where  $\alpha_B$ is the effective recombination coefficient to excited levels of the hydrogen atom, and $h\nu_{H\alpha}$ the energy of an \ha\ photon. The coefficient $\alpha_B$ has a weak dependence on the electron temperature. We adopt the \citet{Draine2011book} parametrization, based on \citet{storey95}, and use our measured $T_e$ -- taken as the average between the low- and high-ionization zones -- along with its variation with galactocentric distance (Sect.~\ref{sec:in-situ-density}), in order to assign an appropriate $\alpha_B$ value to each \hii\ region. 
For a spherical \hii\ region of radius $R$, the previous equation
can be rewritten as

\begin{equation}
\begin{split}
&\nerms\ [\mathrm{cm}^{-3}\mathrm{]} = \\
&1.52\times10^{-16} \sqrt{\frac{L_{\mathrm{H}\alpha}\mathrm{[erg\, s} ^{-1}\mathrm{]} \, (1+y^{+}) \, t_e^{0.833+0.035 \ln t_e}}{R\mathrm{[pc]}^3}} \quad,
\label{eq:ne_rms}
\end{split}
\end{equation}
with $t_e = T_e[K]/10000$. The factor $(1+y^{+})$ accounts for the contribution of ionized helium. We adopt 
$y^{+}=0.08$ as a representative value for the He$^+$/H$^+$ ratio in extragalactic \hii\ regions \citep[][]{Kennicutt1984, Bresolin09, zb12,  Toribio16, Esteban2020}.
Thus, the \hii\ region catalogue allows us to easily derive estimates of \nerms, under the assumption of sphericity of the ionized regions, by means of the equivalent radius ($R_{\mathrm{eq}}$, Sect.~\ref{catalogue_generation}).
It is important to note that \nerms\ is sensitive to the radius that is adopted.

We find that in NGC 2403 the rms electron densities span a range of 1.4–3.4~cm$^{-3}$, with a median value of 
$1.8\pm0.3$~cm$^{-3}$,  whereas in NGC~628 they are somewhat lower, ranging from 0.8 to 2.1~cm$^{-3}$, with a median of $1.1\pm0.2$~cm$^{-3}$.
For the sake of completeness, the median values obtained from luminosities calculated without local background correction are $2.1\pm0.4$ for NGC~2403 and $1.2\pm0.2$~cm$^{-3}$ for NGC~628, which are, within the uncertainties, consistent with those derived using the background-corrected measurements.
\subsection{Electron size-density relation}
\label{sec:ne-R}

\begin{figure*}
    \centering
    \includegraphics[width=0.45\linewidth]{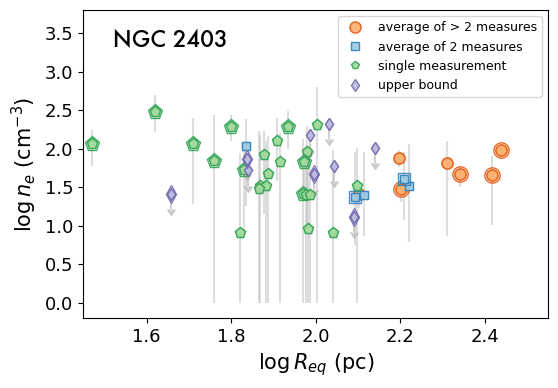}
    \includegraphics[width=0.45\linewidth]{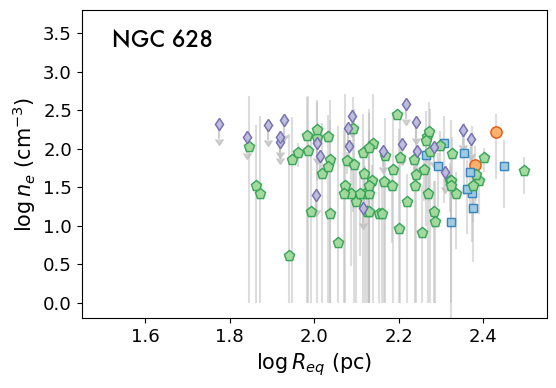}
    \caption{In-situ electron density \ne\ as a function of the equivalent radius (in logarithmic scale) for the \hii\ regions in NGC~2403 (left) and NGC~628 (right). Each blue or orange symbol shows the average of the \ne\ values derived for the same \hii\ region by different authors. The error bars represent the scatter calculated from error propagation (2 independent measurements) or 
    the standard deviation ($>2$ measurements), see Sect.~\ref{sec:in-situ-density}. Single measurements are shown in green (with the corresponding error bars), while upper bounds are indicated by the purple downward arrows. Double symbols for the NGC~2403 regions indicate measurements from our spectroscopic observations (Sect.~\ref{sec:spec}).}
    \label{fig:Ne_vs_Req}
\end{figure*}

Fig.~\ref{fig:Ne_vs_Req} displays the in-situ (\ne) electron density of the \hii\ regions observed in the two galaxies examined in this work, as a function of $R_{\mathrm{eq}}$, in logarithmic units.
In the case of NGC~2403 we can see that the largest \hii\ regions tend to exhibit lower in-situ electron densities (\ne\, $\lesssim 100$ cm$^{-3}$), although this may be the result of the relatively small number of data points available. Such a trend is not observed in NGC~628. The picture is further blurred by the fact that at densities \ne\, $\lesssim 50$ cm$^{-3}$ the diagnostic based on the [\sii] doublet saturates, becoming virtually insensitive to the electron density.
Overall, no clear dependence of $n_e$ on $R_\mathrm{eq}$ is apparent, although the smallest \hii\ regions—particularly in NGC~2403—exhibit a somewhat larger scatter.

\begin{figure*}
\centering
\includegraphics[width=0.45\linewidth]{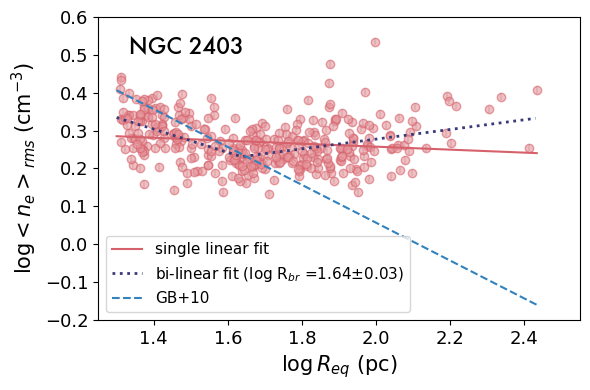}
\includegraphics[width=0.45\linewidth]{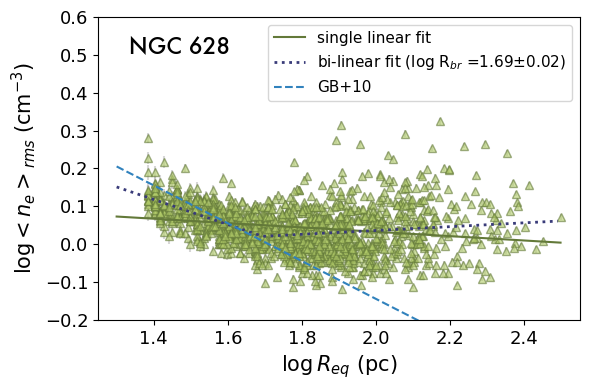}
\caption{Rms electron density \nerms\ as a function of the equivalent radius (in logarithmic scale) for the catalogued \hii\ regions in NGC~2403 (left) and NGC~628 (right).  The solid lines show the best linear fits to all data points, while the dotted lines represent the corresponding bilinear fits, which are statistically favoured over a single linear model ($\Delta$BIC$<-10$; see text for details).  The bilinear fits yield break radii at $\log R_{\mathrm{eq}}\mathrm{[pc]} = 1.64 \pm 0.03$ and $1.69 \pm 0.02$ for NGC~2403 and NGC~628, respectively, with slopes below the break of $-0.31 \pm 0.04$ and $-0.33 \pm 0.02$. The blue dashed line shows the  $\nerms \propto R^{-0.5}$ relation derived by \citet{GB10}. The line has been shifted vertically to match approximately the location of the data points at $\log R_{\mathrm{eq}} \sim 1.7$.}
\label{fig:Ne_rms_Req}
\end{figure*}


In contrast, \nerms, which is one to two orders of magnitude lower ($\sim$0.8-3.4~cm$^{-3}$), exhibits a clear dependence on $R_{\mathrm{eq}}$, as seen in Fig.~\ref{fig:Ne_rms_Req}. A similar trend is found in both galaxies: for \hii\ regions with $R_{\mathrm{eq}}\lesssim60$~pc, $\log \nerms $ decreases approximately linearly with increasing $\log R_{\mathrm{eq}}$.
For larger regions, however, this behaviour breaks down; the scatter increases and $\log \nerms$ shows a slight tendency to rise with  $\log R_{\mathrm{eq}} $. The break radius is consistent with that derived from the $\log L_{\mathrm{H}\alpha}$-$\log R_{\mathrm{eq}}$ relation (Sect.~\ref{sec:catalogues}), as expected given the explicit dependence of $\nerms$  on this relation (Eq.~\ref{eq:ne_rms}). We have nevertheless confirmed this result through an independent bilinear fit to  the $\log \nerms$–$\log R_{\mathrm{eq}}$ relation. A bilinear model is statistically favoured by a significant decrease in the Bayesian information criterion, with $\Delta\text{BIC}$ values for both galaxies falling well below the threshold of $-$10 \citep[e.g.,][]{Liddle2007}. This indicates decisive evidence for the bilinear parametrization over a single linear fit.
 The resulting break values are $\log R_{\mathrm{eq}}\mathrm{[pc]} = 1.64\pm0.03$  for NGC~2403 and $\log R_{\mathrm{eq}}\mathrm{[pc]} = 1.69\pm0.02$ for NGC~628. The $\nerms \propto R_{\mathrm{eq}}^{-\alpha}$ power law observed for the smaller \hii\ regions is in agreement with the results from \citet{GB10} and \citet{Cedres2013} for NGC~4449 and M51, and NGC~6946, respectively. However, these authors obtain $\alpha=0.5\pm0.5$, while our results point to a lower index, $\alpha\simeq0.3$ ($\alpha=0.31\pm0.04$ and $0.33\pm0.02$ for NGC~2403 and NGC 628, respectively).

Both \citet{GB10} and \citet{Cedres2013} applied a correction factor to their $\nerms$ values, in order to account for an observed decrease of $\nerms$ with increasing galactocentric radius. Our data include no such correction, as we find no evidence for such a trend neither in NGC~2403 nor NGC~628. However, we note that the electron densities derived from luminosities uncorrected for local DIG emission show a decrease with galactocentric radius in NGC~2403. This behaviour arises because the H$\alpha$ surface brightness of the DIG typically declines with radius \citep[e.g.][]{Zurita2000}. As discussed in Sect.~\ref{DIG_correction}, luminosities (and therefore densities) obtained without applying a DIG correction represent upper limits to the true values. We cannot rule out an intrinsic dependence of \nerms\ on galactocentric radius, which would in fact be expected if the \hii\ regions are in quasi–pressure equilibrium with their surroundings, given the decline of the ambient ISM pressure with radius \citep[e.g.][]{Elmegreen1989,BlitzRosolowsky2004}. Nevertheless, the radial variation that we obtain for NGC~2403 from uncorrected luminosities is much shallower than that found by \citet{GB10} for NGC~4449 and M51, with scale lengths of $\sim$98~kpc (cf. $\sim$10–11~kpc in NGC~4449 and M51; \citealt{Cedres2013} report no value). Likewise, \ne\ shows no clear dependence on galactocentric radius in NGC~2403  and NGC~628. 

\begin{figure*}
    \centering
    \includegraphics[width=0.45\linewidth]{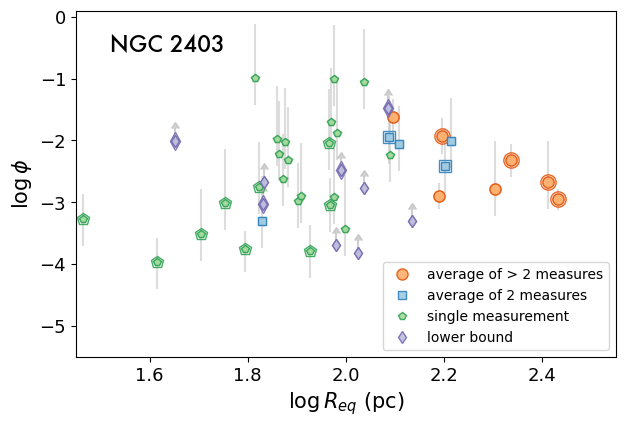}
    \includegraphics[width=0.45\linewidth]{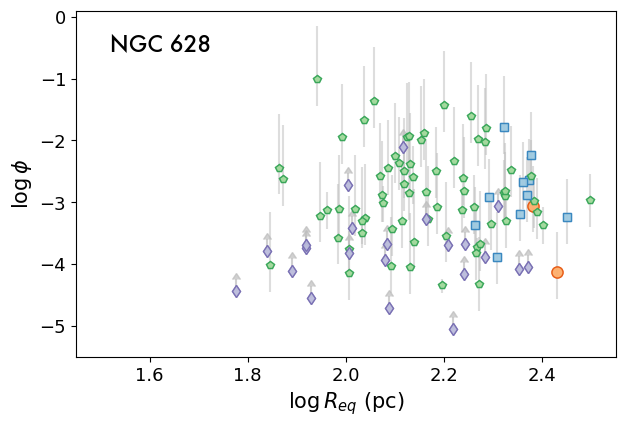}
    \caption{Logarithm of the filling factor $\phi$ as a function of the logarithm of the \hii\ region equivalent radius R$_{eq}$ for \hii\ the regions in NGC~2403 (left) and NGC~628 (right) for which we could make an estimate of the two electron densities (in-situ and rms). Error bars were derived by propagating the uncertainties in \ne\ and \nerms.}
    \label{fig:FF}
\end{figure*}
\subsection{\texorpdfstring{Filling factor, $\phi$}{Filling Factor}}
It is well established that electron densities derived from the emission line spectroscopy of \hii\ regions are significantly higher than those inferred from their observed luminosities  (e.g. \citealt{Kennicutt1984, Martin1997}; \citetalias{HH09}). This discrepancy is clearly manifest in our data (comparing Figs.~\ref{fig:Ne_vs_Req} and \ref{fig:Ne_rms_Req}). 
It is commonly attributed to the presence of density inhomogeneities or fluctuations within the regions (see Sect.~\ref{sec:intro}), and is usually quantified through the {\em filling factor}, $\phi$, defined as the fraction of the \hii\ region volume occupied by high–density clumps of uniform density equal to \ne, $\phi=(\nerms/\ne)^2$.  Fig.~\ref{fig:FF} shows the filling factor as a function of the equivalent radius, in logarithmic scale, for the \hii\ regions for which both electron density estimates are available in NGC~2403 and NGC~628. For these regions, since the value of {\it c}(\hb)  is available, $\phi$ was computed using extinction–corrected \ha\ luminosities. All regions exhibit filling factors below $10^{-1}$, with most lying in the range $10^{-4}$ – $10^{-2}$, in overall agreement with previous determinations for extragalactic \hii\ regions \citep{KennicuttHodge80,Kennicutt1984, Rozas1996, n7479_hii, Cedres2013}.

In apparent contrast to the results of \citet{Cedres2013} for NGC~6946, our data show no dependence of $\phi$ on either the radius or the luminosity of the \hii\ regions. 
However, we cannot rule out the presence of such relations in our galaxies. Our estimates of $\phi$ are limited to the largest and most luminous regions ($\log R_{\mathrm{eq}}\gtrsim$1.6–2.5), namely those for which spectroscopic observations allow the determination of \ne. \citet{Cedres2013} sample a broader range in radius, extending to smaller sizes ($\log R_{\mathrm{eq}}\gtrsim$1–2.3), 
but within the same size range covered by our data it is difficult to
ascertain even from their observations that a clear dependence exists,
due to the large scatter in the values of $\phi$ they report.
Thus, while a size dependence cannot be excluded for the \hii\ regions in NGC~628 and NGC~2403, our current data do not allow us to detect it.


\section{Discussion}
\label{sec:discussion}
The negative correlation between the rms  electron density, \nerms, and the diameter, $D$, of \hii\ regions was first identified in Galactic sources \citep{HabingIsrael1979}. Subsequent work showed that extragalactic \hii\ regions occupy the low–density, large–diameter extension of this relation  \citep{Kennicutt1984}. Over the following decades, a power-law dependence of the form $\nerms \propto D^{-1}$ has been consistently confirmed  through observations of compact and ultracompact \hii\ regions \citep[e.g.][]{Garay1993, GarayLizano1999, KimKoo2001, MartinHernandez2003}. 
In order to investigate the physical origin of this relationship, \citetalias{HH09} compiled both Galactic and extragalactic samples spanning six orders of magnitude in density and size. They demonstrated that neither static nor evolutionary models of \hii\ regions, including pressure-driven expansion \citep{HH06},  could explain the observed relation for the full dataset with a single set of initial conditions. 
Instead, the size-density relation  is best interpreted as a sequence containing objects with vastly different initial gas densities, suggesting that star formation processes preserve the imprint of the parental molecular environment. For compact and ultracompact \hii\ regions, for which $\nerms \gtrsim 10^3$~cm$^{-3}$, dust extinction was also found to play a key role in limiting their sizes.

\begin{figure*}
    \centering
    \includegraphics[width=0.6\linewidth]{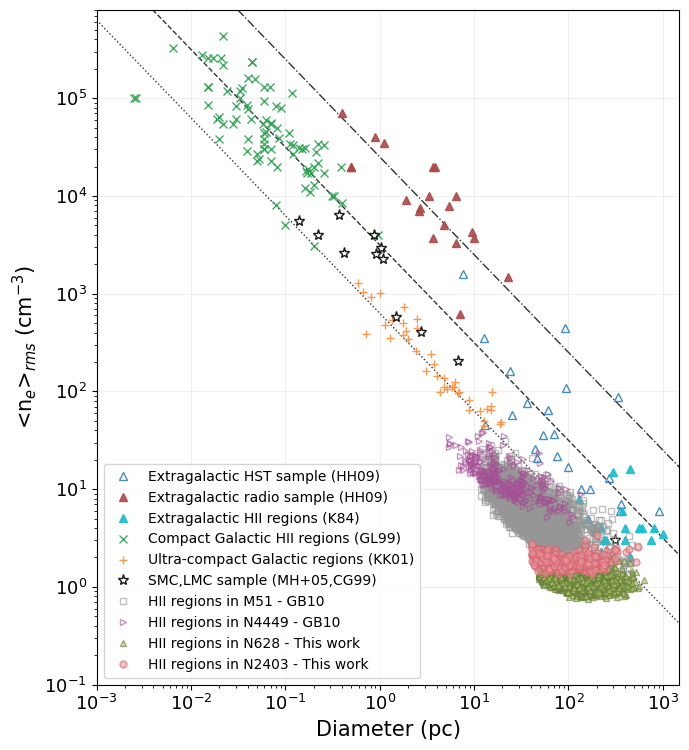}
    \caption{Updated version of the \citetalias{HH09} size–rms density plot for Galactic and extragalactic \hii\  regions \citep[including data from][]{Kennicutt1984,GarayLizano1999,CG99,KimKoo2001,MH2005}, now including the \hii\ regions in NGC~2403 and NGC~628 catalogued in this work, together with those from \citet{GB10} for NGC~4449 and M51. Straight lines, as in the original \citetalias{HH09} plot, indicate unit–slope relations with intercepts ($\log \nerms$~[cm$^{-3}$] at a diameter of 1 pc) of 2.8, 3.5, and 4.4.}
    \label{HH09plot}
\end{figure*}

Although the number of extragalactic studies on this topic is still quite limited, some authors have found densities and sizes of \hii\ regions in nearby galaxies that are consistent with those studied by \citet{Kennicutt1984}. This is the case of the data  analysed by \citet{GB10} on M51 and NGC~4449, by \citet{Cedres2013} on NGC~6946, by \citet{Maragkoudakis2018} for about 20 \hii\ regions in M33 and M83 and, more recently, by \citet{Boselli2025} for Virgo Cluster galaxies.  Considering the full population of \hii\ regions within a single galaxy, the size-density relation is also found to follow a power law, $\langle n_{\mathrm{e}} \rangle_{\mathrm{rms}} \propto D^{-\alpha}$, although with a relatively shallow slope. Previous studies (\citealt{GB10,Cedres2013}) report $\alpha = 0.50\pm 0.05$, where the uncertainty represents galaxy-to-galaxy variations.  Our pilot study on NGC~2403 and NGC~628  also reveals a size-density anti-correlation, with a slope shallower than unity (Fig.~\ref{fig:Ne_rms_Req}), in agreement with these authors. However, our results differ in three  aspects: 

\begin{enumerate}[label=\alph*), labelsep=0.5em, leftmargin=*, itemsep=3pt]
\item we obtain an even flatter slope ($\alpha\simeq0.3$)
\item the size--density anti-correlation is only apparent for the smallest and faintest \hii\ regions, showing a  break at an apparently fixed effective radius  $R_{\mathrm{eq}}\approx$45--50~pc
\item our sample of extragalactic \hii\ regions departs from the \citetalias{HH09} sequence, with most objects offset toward smaller diameters (see Fig.~\ref{HH09plot}).
\end{enumerate}

The very limited number of galaxies so far analysed makes it difficult to determine the origin of a). However, given the multiple tests performed to asses the reliability of our catalogues, we have strong reasons to believe that, although genuine galaxy-to-galaxy variations in the slope cannot be ruled out, the adopted methodology and/or the image resolution may significantly affect the derived slope of the size–density relation in a given galaxy. In particular,  measuring  fluxes from the higher–resolution (non–degraded) \ha\ images while using segmentation maps derived from the degraded one (see Sect.~\ref{sec:cat_method}), can produce slopes $\alpha\approx0.45$–0.50, in good agreement with those reported by \citet{GB10} and \citet{Cedres2013}. In both these studies, the physical resolution of the \ha\ images is smaller than ours, supporting this interpretation.  Nevertheless, a larger sample of galaxies is required to test whether the slope also depends on intrinsic galaxy properties.

The break in the relation mentioned in b) is a persistent feature, consistently appearing in all the tests performed, and always occurring at a characteristic size of $\log R_{\mathrm{eq}}[\mathrm{pc}] \simeq 1.6$–$1.7$. A similar transition was also identified by \citet{Zaragoza-Cardiel2013, Zaragoza-Cardiel2014} in the \hii\ regions of  interacting systems, a feature that appears to be less frequent  in their sample of isolated galaxies \citep{Zaragoza-Cardiel2015}. They interpreted the break as arising from the coexistence of two distinct populations of \hii\ regions, the smaller ones being pressure-confined, while the larger and more luminous regions being gravitationally bound by their own self-gravity. We note, however, that recently \citet{Pathak2025}, studying a large number of \hii\ regions in 19 galaxies, concluded that self-gravity plays a secondary role compared to ambient ISM pressure in regulating their dynamics.
Following \citet{Zaragoza-Cardiel2015}, we assume that \hii\ regions with sizes below the break radius $R_\mathrm{br}$ are pressure-confined by the surrounding ISM. In this scenario, the observed radius corresponds to the equilibrium radius at which the internal pressure of the region balances the external (ambient) pressure. Considering that the ionizing clusters of \hii\ regions are approximately located in the mid-plane of the galactic disc, the observed correlation between \nerms\ and the equilibrium radius $R_\mathrm{eq}$ (for $R < R_\mathrm{br}$) can be interpreted as a tracer of how the ambient ISM density (or pressure) varies with height $z$ above the galactic plane. Fitting to our data in Fig.~\ref{fig:Ne_rms_Req} the relation $\nerms\ = {\langle}n_e{\rangle}_{\mathrm{rms},0}  \, e^{-R_{\mathrm{eq}}/h_z}$,  where $R_\mathrm{eq}$ serves as a proxy for $|z|$, we obtain scale heights of $h_z = (100 \pm 13$)~pc for NGC~2403 and $h_z = (109 \pm 6)$~pc for NGC~628. These values are consistent with the scale height measured for the Milky Way thin disc \citep[$\sim$100-140~pc for the ionized and cold neutral medium --][]{DickeyLockman1990, CordesLazio2002, KalberlaKerp2009}, lending support to this interpretation.

Fig.~\ref{HH09plot} presents an updated version of the \citetalias{HH09} size–density diagram, now including the \hii\ regions in NGC~2403 and NGC~628 catalogued in this work, as well as those in NGC~4449 and M51 from \citet{GB10}, for which we have re-computed \nerms\ following Eq.~\ref{eq:ne_rms} for $T_e$=$10^{4}$~K. The latter two galaxies are the only systems we know for which the \nerms\ of their \hii\ region populations has been analysed using a methodology similar to ours, with publicly available data. Fig.~\ref{HH09plot} clearly illustrates the behaviour described in c): the bulk of the \hii\ regions in these galaxies appears systematically offset towards  smaller diameters relative to the prediction by the \citetalias{HH09} relation. The original version of this diagram incorporated the 24 extragalactic \hii\ regions compiled by \citet[]{Kennicutt1984}. Importantly, this sample includes only a few of the most luminous and extended regions in nearby galaxies, namely LMC, SMC, M33, M31, IC~1613, M81, and M101. Looking at the location of these photoionized nebulae in the size-density diagram, we find good agreement with the largest (and  brightest) \hii\ regions in our sample and in the \citet{GB10} sample. Therefore, although the size–density relation for \hii\ regions, parametrized as $\nerms \propto D^{-\alpha}$, is well known and widely used, our results indicate that caution is required in practical applications. The inclusion of more extensive samples of extragalactic \hii\ regions, as done here, increases the dispersion and reveals that, over the relatively narrow range in size and density typical of these nebulae, the relation cannot be easily described by an exponent $\alpha=1$.

\begin{figure}
    \centering
    \includegraphics[width=\linewidth]{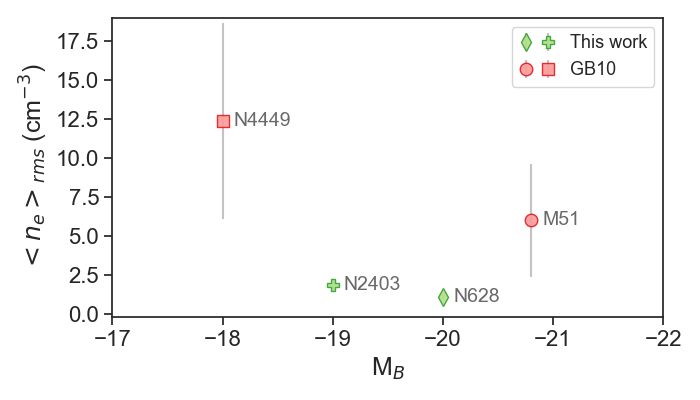}
    \caption{Median value of the rms electron density of the \hii\ regions of different galaxies as a function of their absolute $B$-band magnitude.}
    \label{fig:ne_vs_M}
\end{figure}

It is worth noting that the median value of \nerms\ for the \hii\ regions in NGC~2403 and NGC~628 differs, being slightly higher in NGC~2403 ($1.8\pm0.3$~cm$^{-3}$) than in NGC~628 ($1.1\pm0.2$~cm$^{-3}$). We verified that this difference persists even when accounting for methodological factors that could affect the derived densities, such as local background subtraction or variations in the catalogue extraction parameters. Such potential inter-galaxy differences appear even more pronounced considering M51 and NGC~4449. Although not explicitly discussed in the original publication by \citet{GB10}, an analysis of their published catalogues yields median values of $6\pm4$~cm$^{-3}$ and $12\pm6$~cm$^{-3}$, respectively, with the error representing the rms dispersion of the data values. Fig.~\ref{fig:ne_vs_M} shows, as a function of galaxy absolute $B$ magnitude, the median value of \nerms\ for our sample galaxies, together with those derived from the \citet{GB10} study.  A large scatter is evident, with no obvious correlation with the luminosity (or, equivalently, mass) of the host galaxies. 
The in-situ electron density \ne\ (not shown) appears to be much more uniform from galaxy to galaxy than the rms density. The observed variations in \nerms\ hint at differences in the filling factor or porosity of the ionized gas.  This would manifest observationally in the fact that, for a given ionizing source, \hii\ regions would appear systematically larger in galaxies with lower median filling factors or \nerms. 

\begin{figure}
    \centering
    \includegraphics[width=\linewidth]{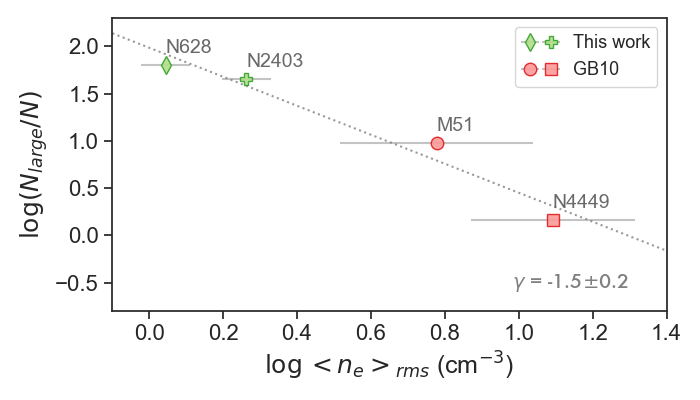}
    \centering
    \includegraphics[width=\linewidth]{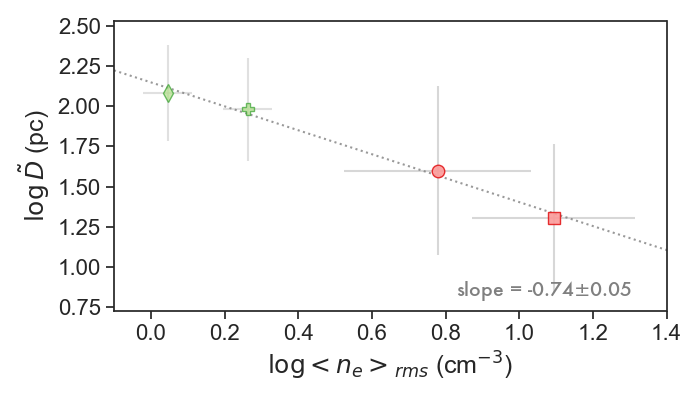}
    \centering
    \includegraphics[width=\linewidth]{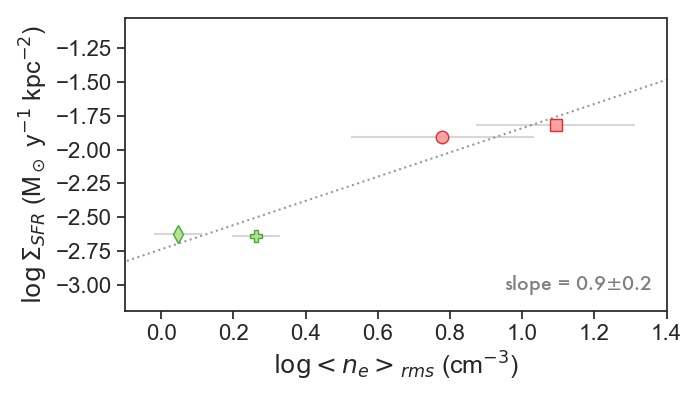}
    \caption{{\em Top}: logarithm of the fraction of large \hii\ regions ($R_{\rm eq} > 50$~pc); {\em middle:} logarithm of the median \hii\ region diameter; {\em bottom:} logarithm of the integrated SFR surface density. All quantities are plotted as a function of the logarithm of the median $\nerms$ of the \hii\ regions in NGC~2403 and NGC~628 (this work), and in NGC~4449 and M51 from \citet{GB10}. The straight lines represent the best linear fits, with the corresponding slopes indicated in each panel. Error bars show the standard deviation of the $\nerms$ or diameter distributions. The $\Sigma_{\rm SFR}$ values in the bottom panel are derived using measurements of \citet{Leroy2008}.}
    \label{fig:props_vs_ne}
\end{figure}

Differences in the diameter  distributions  of  \hii\ regions between galaxies have been recognized since the 1980's. It is well established that, as first discussed by \citet{van_den_Bergh1981}, the cumulative distribution of \hii\ region diameters in a spiral or irregular galaxy follows an exponential law of the form $N(>D)=N_0 \exp{(-D/D_0)}$, where $D_0$, the characteristic diameter, scales linearly with the absolute magnitude $M_B$ of the host galaxy \citep{Hodge1983,Hodge1987}, with less luminous (and less massive) galaxies having a  smaller characteristic diameter. This trend has been confirmed in subsequent studies \citep[e.g.][]{Ye1992, Rozas1996, Knapen1998, deSouza2018, DellaBruna2022}, yet its underlying physical origin remains unexplained. Differences in the mean rms electron density among galaxies provide a plausible explanation. \citet{Hodge1983, Hodge1987} speculated that the $D_0 -  M_B$ relation could be linked to the gas density, suggesting that the relative number of large \hii\ regions in a galaxy could depend on some power of the gas density, while the gas density itself depends more weakly on the galaxy mass or luminosity. 
Assuming that the electron density of \hii\ regions is determined by the gas density of their parental molecular clouds (\citetalias{HH09}; \citealt{Davies2021}), we can directly test Hodge's hypothesis. 
The weak dependence between \nerms\ and $M_B$,  already  shown in Fig.~\ref{fig:ne_vs_M}, lends support to this interpretation. The top panel of Fig.~\ref{fig:props_vs_ne}\footnote{We verified that the relations shown in this figure remain unchanged—yielding slopes consistent within the uncertainties—when applying an extinction correction to \nerms\  based on the average \hii\ region extinction in the four galaxies, using the values measured here for NGC~2403, those compiled by \citet{PaperI} for NGC~2403, NGC~628, and M51, and the mean value reported by \citet{QuillenYukita2001} for NGC~4449.} shows  the relative number of large \hii\ regions in each catalogue as a function of the median value of \nerms, on a logarithmic scale. As a working definition for “large” \hii\ regions, we include those with radii greater than 50~pc, corresponding approximately to the break observed in the $\log~L_{\mathrm{H}\alpha}-\log~R_{\mathrm{eq}}$ or $\log~\nerms-\log~R_{\mathrm{eq}}$ relations (Sects.~\ref{sec:catalogues} and \ref{sec:ne-R}). Both our data and those from \citet{GB10} are included. Although caution is required when interpreting these results -- they are based on only four galaxies analysed with slightly different methodologies (see Sect.~\ref{sec:appenN628}) -- the data indicate that the relative number of large \hii\ regions  may decrease with increasing \nerms, following a power law of the form: 
\begin{equation}
\frac{N_{\mathrm{large}}}{N} \propto \nerms^{-\gamma} 
\end{equation}

\noindent with $\gamma=1.5\pm0.2$. We verified that normalizing the number of regions by $N_0$, or changing the definition of “large” \hii\ regions, alters the slope of the power-law fit within the range $-2.0$ to $-1.4$. 

In order to further understand and interpret these scaling relations, we assume that the majority of the relatively large ($R_{\mathrm{eq}}>25$ pc) \hii\ regions we observed have reached pressure equilibrium with the surrounding cold ISM (\citealt{OeyClarke1997, Elmegreen2000, GB10, Shirazi2014}), while smaller \hii\ regions are likely overpressured (\citealt{Pathak2025}).  Under our assumption, the rms density scales with the pressure of the ambient ISM (\citealt{Elmegreen2000}).  In the dusty \hii\ region models by \citet{Draine2011}, constant confining pressure locations in the size-density diagram correspond to virtually constant rms densities, within the parameter regime occupied by the extragalactic \hii\ regions we study here. This author also pointed out that the $-1$ slope of the relation studied by \citetalias{HH09} (and previous authors) is possibly an artefact of the sample selection. We agree with this conclusion, at least in the case of extragalactic \hii\ regions, as these objects  cover a range of sizes and densities that are regulated by variations of intrinsic parameters, such as the ionizing luminosity and the ambient pressure. The fact that the \hii\ regions we have studied in NGC~2403 and NGC~628 do not follow the "traditional" $-1$ slope of the relation \citepalias{HH09} is in line with this reasoning.

Focusing, in order, on NGC~4449, M51, NGC~2403, and NGC~628 in Fig.~\ref{HH09plot}, we notice a clear progression to smaller mean rms densities and larger mean diameters of their \hii\ regions. This is also illustrated in the middle panel of Fig.~\ref{fig:props_vs_ne}, where the median \hii\ region diameter of these four galaxies is plotted against their median \nerms, in logarithmic units.  A linear fit indicates a slope of $-0.74\pm0.05$ (but we are cautious about generalizing this result, based on only four galaxies).  Different galaxies appear to have different mean rms densities, and correspondingly different mean ambient ISM pressure values. A larger mean ambient pressure determines smaller mean \hii\ region sizes, as expected. This can also help us interpret the diagram in the top panel of Fig.~\ref{fig:props_vs_ne}, introduced earlier: lowering the ISM pressure facilitates the creation of relatively large \hii\ regions. As mentioned above, in its pioneering study of the size distribution of \hii\ regions in irregular and spiral galaxies \citet{Hodge1983, Hodge1987} proposed that the relative number of large \hii\ regions could depend on the gas density. We believe that our interpretation justifies Hodge's statement. Furthermore, if confirmed with a larger sample, the relation in Fig.~\ref{fig:props_vs_ne} (top) could provide valuable input for models describing the formation of massive bound stellar clusters. Since the largest \hii\ regions are also the most luminous (Fig.~\ref{L-V}), the figure yields the fraction of massive ionizing stellar clusters as a function of the median rms density, given the observed proportionality between \ha\ luminosity and ionizing cluster mass \citep{Scheuermann2023}. This observational link may then help constrain the cluster mass function and its dependence on environmental conditions—such as ambient pressure—as well as on global galaxy properties including the star formation rate (SFR) and associated feedback processes \citep[e.g.][]{Elmegreen2008,Krumholz2019,Adamo2020,Andersson2024}. 

Lastly, we show in the bottom panel of Fig.~\ref{fig:props_vs_ne} how the rms electron density correlates with the surface density of the star formation rate, \SigmaSFR, calculated from the star formation rates and isophotal radii presented in \citet{Leroy2008}. The linear fit yields \SigmaSFR $\propto$ $\nerms^{0.9\pm0.2}$. A number of recent studies at high-redshift, up to $z\sim10$, have shown that the in-situ electron density, \ne, of star-forming galaxies, measured from their integrated spectra, increases with redshift (e.g. \citealt{Shimakawa2015, Kaasinen2017, Davies2021, Topping2025, Li2025}). At the same time, a relation between \SigmaSFR\ and \ne\ is also found. The observed increase of the ionization parameter $U \propto (Q n_e \phi^2)^{1/3}$ with redshift at constant stellar mass has been linked to a significant increase in the ionizing photon rate $Q$ (and thus star formation rate) and electron density (\citealt{Reddy2023a, Reddy2023}), since the volume filling factor $\phi$ does not appear to evolve substantially (\citealt{Davies2021}).

Local, low-redshift results in support of the increased electron density with increased star formation activity and/or \SigmaSFR\ have been presented, among others, by \citet{Brinchmann2008}, \citet{Liu2008}, \citet{Bian2016}, \citet{Herrera-Camus2016} and \citet{Kaasinen2017}. \citet{Jiang2019} found that \SigmaSFR\ of local green pea galaxies and Lyman-break analogues increases with thermal pressure as $P_{th}^{1.33}$ (\SigmaSFR$\propto$ $n_e^{1.33}$, since $P_{th}\simeq2 n_e k T$). \citet{Li2025} find an index of 2.0 for a sample of low redshift ($z<0.04$) systems. For galaxies at redshifts between 1.9 and 3.7, \citet{Reddy2023a} find \SigmaSFR$\propto$ $n_e^{2.0}$ (compared to an index of $\sim$1.6 from \citealt{Shimakawa2015}). We can thus conclude that our findings (relative to \nerms\ rather than $n_e$) are in rough agreement with these results, albeit being characterized by a flatter slope (which is based on a very small number of data points). We also note that \citet{Li2025} find that $n_e$ remains constant below a certain threshold, $\log \SigmaSFR$~[M$_\odot$~yr$^{-1}$~kpc$^{-2}$] $< -1.46$, where all our data points are found (see Fig.~\ref{fig:props_vs_ne}, bottom).

The explanation for the correlation between gas density and level of star formation potentially lies in an interplay between factors, that include the density of the parent molecular clouds out of which stars form, the dependence of star formation activity on the surface density of the cold gas (Kennicutt-Schmidt law), the effects of stellar feedback on \hii\ region evolution, the dynamics of the ISM and the pressure balance in the expanding \hii\ regions (see discussions in \citealt{Herrera-Camus2016, Davies2021, Li2025}).

We stress that virtually all the studies we mentioned have focused on the in-situ (surface brightness-weighted) electron density, as measured from optical doublet line ratios ([\oii] $\lambda3726/\lambda3729$ and [\sii] $\lambda6716/\lambda6731$) observed in the integrated spectra of star-forming galaxies.  We have instead centred our analysis on the rms (volume averaged) electron density,  which, unlike in the studies mentioned above, has been derived from the \ha\ luminosities and sizes of the spatially resolved individual \hii\ regions.
In their study of both distant and nearby star-forming galaxies, \citet{Davies2021} have derived both $n_e$ and \nerms\ (actually a lower limit to this quantity, since \hii\ regions do not fill the galaxy volumes), finding that their evolution with redshift is coupled, and thus that the filling factor remains virtually constant through the range $0 < z < 2.6$, with a value on the order of $10^{-3}$ (comparable to what we measure in NGC~2403 and NGC~628 -- see Fig.~\ref{fig:FF}). As these authors explain, this finding is essential in order to link the (in-situ) density variations with redshift, observed for star-forming galaxies, to physical processes and gas conditions.
The fact that the level of star formation in galaxies is fundamentally tied to the volume-averaged, rms electron density, highlights the importance of measuring \nerms\ in larger samples of \hii\ regions and galaxies.

\section{Conclusions}
\label{sec:conclusions}
We have analysed the root-mean-square electron densities (\nerms), in-situ electron  densities (\ne), and volume filling factors ($\phi$) of \hii\ regions in the nearby spiral galaxies NGC~2403 and NGC~628. This work represents the first step of a broader project aimed at understanding how these quantities depend on both the intrinsic properties of \hii\ regions and the physical conditions of their host galaxies, with the ultimate goal of identifying the factors that regulate the porosity of the interstellar medium and, consequently, the escape of ionizing radiation.

The derivation of  \nerms\ requires measurements of the \ha\ luminosity and of  the emitting volume, or the equivalent radius $R_{\text{eq}}$, assuming spherical \hii\ regions. The latter parameter is particularly sensitive to the methodology and identification criteria adopted. Our comparison with published catalogues for NGC 628 highlights the need for a homogeneous approach, which is  essential for meaningful comparisons between galaxies. To this end, we have developed a new, uniform methodology based on image segmentation to construct comprehensive \hii\ region catalogues from \ha\ imaging, which has been applied here to NGC~2403 and NGC~628 as pilot galaxies. These catalogues comprise 1458 and 370 regions, respectively. In addition, \ne\ was determined from the [\sii]$\lambda\lambda6717,6731$ doublet for subsamples of 36 regions in NGC~2403 (plus 9 upper limits) and 80 regions in NGC~628 (plus 22 upper limits), using a combination of new and compiled literature fluxes, allowing for the determination of $\phi$.  
Our main results can be summarized as follows:

\begin{itemize}[labelsep=0.5em, leftmargin=1em, itemsep=1pt]


\item The \ne\ values are typically below 300~cm$^{-3}$, with a median value of approximately 45~cm$^{-3}$  for both galaxies. No clear dependence on region size or galactocentric radius is observed.

\item The \nerms\ densities are one to two orders of magnitude lower than \ne, with median values of $1.8\pm0.3$~cm$^{-3}$ NGC~2403 and $1.1\pm0.2$~cm$^{-3}$ for NGC~628. 

\item Assuming a two-phase model for the \hii\ regions, we derive volume filling factors, $\phi$, that range from $\sim10^{-4}$ to $10^{-1}$, in agreement with values reported in the literature for extragalactic \hii\ regions. No significant dependence on region properties or galactocentric radius is found.

\item The bulk of the nebulae we studied are systematically offset toward smaller diameters relative to the well-known sequence analysed by \citetalias{HH09}, indicating that the traditional  $\nerms \propto D^{-1}$ relation does not accurately describe the trend observed in giant extragalactic \hii\ regions within a single galaxy.

\item Both our target galaxies exhibit a similar size–density relation $\nerms \propto R_{\rm eq}^{-\alpha}$  for  $R_{\rm eq}\lesssim 50$~pc, with an index $\alpha \sim 0.3$. This anti-correlation breaks for regions larger than $\sim50$~pc.

\item Combining our results with published measurements for two additional galaxies \citep[NGC 4449 and M51 from][]{GB10} suggests interesting tentative scaling relations with galactic properties:

\begin{itemize}
\item The fraction of large \hii\ regions ($R_{\mathrm{eq}} \gtrsim 50$~pc) may decrease with increasing median \nerms, following a power law: $N_{\mathrm{large}}/N \propto \nerms^{-\gamma}$ , with  $\gamma=-1.5\pm0.2$, supporting an early hypothesis by \citet{Hodge1983, Hodge1987} linking gas density to the relative number of large \hii\ regions.

\item A higher median \nerms\ -- possibly tracing a larger ambient pressure -- correlates with smaller median \hii\ region sizes.

\item The median \nerms\ correlates with the SFR surface density: \SigmaSFR $\propto$ $\nerms^{0.9\pm0.2}$.
\end{itemize}

Although these relations should be interpreted with caution, being based on only four galaxies and heterogeneous datasets, they appear promising. Their physical origin is likely rooted in the interplay between the density of the parent molecular clouds, the ambient ISM pressure, and the star-formation activity.
\end{itemize}

Our results provide new constraints for models of massive cluster formation and are important for interpreting observed trends at high-redshift. Extending this analysis to larger, uniformly studied samples of \hii\ regions and galaxies will be essential to fully test and refine these tentative scaling relations.


\section*{Acknowledgements} 
We thank the anonymous referee for constructive suggestions that improved the clarity of the manuscript. We acknowledge support from project PID2023-150178NB-I00 (and PID2020-114414GB-I00), financed by MCIU/AEI/10.13039/501100011033, and by FEDER, UE. We also acknowledge support via grant FQM108, financed by the Junta de Andalucía (Spain). SV and EF acknowledge support from project PID2023-149578NB-I00 financed by MCIU/AEI/10.13039/501100011033, and by FEDER, UE.

AZ is grateful for the kind hospitality received at the Institute for Astronomy at the University of Hawaii in Manoa, Honolulu. FB thanks the Departamento de F\'\i sica Teórica y del Cosmos, Universidad de Granada, for the warm hospitality. Based on observations collected at the Centro Astronómico Hispano en Andalucía (CAHA) at Calar Alto, operated jointly by Junta de Andalucía and Consejo Superior de Investigaciones Cientíﬁcas (IAA-CSIC). The William Herschel Telescope operated on the island of La Palma by the Isaac Newton Group of Telescopes in the Spanish Observatorio del Roque de los Muchachos of the Instituto de Astrofísica de Canarias.
This research made use of astropy, a community-developed core python (http://www.python.org) package for Astronomy \citep{astropy:2013, astropy:2018, astropy:2022}; ipython \citep{ipython_Perez2007}; matplotlib \citep{matplotlib_hunter2007}; numpy \citep{numpy_harris2020}; scipy \citep{scipy_virtanen2020}; and {\sc TOPCAT} \citep{TOPCAT2005}. 

\section*{Data Availability}


The data underlying this article,  the \hii\ region catalogues for NGC~2403 and NGC~628, and the emission-line fluxes of \hii\ regions in NGC~2403, are available in the online supplementary material of the article and on the CDS VizieR facility (\href{https://vizier.u-strasbg.fr/viz-bin/VizieR}{https://vizier.u-strasbg.fr/viz-bin/VizieR}).


\bibliographystyle{mnras}
\bibliography{referencias} 



\appendix

\section{Emission line fluxes}
Tables~\ref{line_fluxes} and \ref{line_fluxes2} contain the emission line fluxes -- normalized to \hb\,=\,100 -- of the newly observed \hii\ regions in NGC~2403, corrected for interstellar reddening, as described in Sect.~\ref{sec:spec}. For each individual \hii\ region the bottom four rows report the extinction coefficient {\it c}(\hb), the extinction-corrected \hb\ flux, the value assumed for the equivalent width of the underlying stellar absorption (EW$_\mathrm{abs}$) in the Balmer lines, and the size of the aperture used to extract the 1D spectra.

\begin{table*}
\begin{threeparttable}
\caption{\label{line_fluxes} Extinction-corrected emission line ratios in the spectra of the new spectroscopic observations of \hii\ regions in NGC~2403 (Sect.~\ref{sec:spec}), normalized to  $\hb = 100$.}
\centering
\begin{tabular*}{\textwidth}{@{\extracolsep{\fill}}lccccccccc}
\hline\hline
Line ID & \multicolumn{9}{c}{\hii\ region} \\
       & s1.1 & s1.2 & s1.3 & s1.4 & s1.5 & s1.6 & s1.7 & s2.1 & s2.2 \\
\hline
[\oii]~$\lambda3727$     & $240\pm40$ & $310\pm10$ & $700\pm100$ & $320\pm10$ & $420\pm70$ & $270\pm40$ & $370\pm60$ & $400\pm20$ & $380\pm30$ \\ 
H$\delta$~$\lambda$4102  & - & $22\pm4$   & - & $28\pm3$ & - & - & - & $30\pm4$ & - \\ 
H$\gamma$~$\lambda$4340  & - & $47\pm3$   & - & $47\pm3$ & - & $33\pm9$ & - & $47\pm3$ & $46\pm7$ \\ 
{}[\oiii]~$\lambda4363$    & - & - & - & - & - & - & - & - & - \\
H$\beta$~$\lambda4861$   & $100\pm10$ & $100\pm4$  & $100\pm20$ & $100\pm3$ & $100\pm9$ & $100\pm9$ & $100\pm9$ & $100\pm4$ & $100\pm6$ \\
{}[\oiii]~$\lambda4959$     & $130\pm10$ & $69\pm3$   & $70\pm20$ & $63\pm2$ & - & $93\pm8$ & $25\pm6$ & $59\pm3$ & $47\pm4$\\
{}[\oiii]~$\lambda5007$ & $400\pm30$ & $210\pm7$ & $100\pm20$ & $178\pm5$ & $63\pm7$ & $280\pm20$ & $68\pm8$ & $158\pm6$ & $105\pm7$\\
{}[\nii]~$\lambda5755$ & - & - & - & - & - & - & - & - & - \\
\hei~$\lambda$5876 & $12\pm4$ & $13\pm1$ & - & $13.3\pm0.8$ & - & $18\pm4$ & $8\pm3$ & $11.1\pm1.0$ & - \\
{}[\oi]~$\lambda6300$ & - & - & - & $6.9\pm0.6$ & - & - & - & - & - \\
{}[\siii]~$\lambda6312$ & - & - & - & - & - & - & - & - & - \\
{}[\oi]~$\lambda6364$ & - & - & - & $2.8\pm0.5$ & - & - & - & - & - \\
{}[\nii]~$\lambda6548$ & - & $11.2\pm0.7$ & $19\pm5$ & $13.5\pm0.7$ & $22\pm4$ & $9\pm3$ & $11\pm3$ & $16\pm1$ & $14\pm2$\\
\ha~$\lambda$6563 & $290\pm20$ & $290\pm10$ & $280\pm50$ & $290\pm10$ & $300\pm50$ & $280\pm50$ & $280\pm50$ & $290\pm10$ & $290\pm10$ \\
{}[\nii]~$\lambda6583$ & $33\pm4$ & $40\pm2$ & $70\pm10$ & $41\pm2$ & $47\pm8$ & $41\pm7$ & $36\pm6$ & $44\pm2$ & $47\pm3$ \\
\hei~$\lambda$6678 & - & $3.1\pm0.5$ & - & $4.0\pm0.4$ & - & - & - & $5.9\pm0.9$ & - \\
{}[\sii]~$\lambda6717$ & $64\pm6$ & $36\pm2$ & $60\pm10$ & $40\pm2$ & $50\pm9$ & $43\pm8$ & $41\pm7$ & $48\pm2$ & $46\pm3$ \\
{}[\sii]~$\lambda6731$ & $41\pm4$ & $26\pm1$ & $50\pm10$ & $28\pm1$ & $41\pm7$ & $34\pm6$ & $30\pm5$ & $32\pm2$ & $32\pm2$ \\
\hei~$\lambda$7065 & - & $2.4\pm0.7$ & - & $2.5\pm0.5$ & - & - & - & - & - \\
{}[\ariii]~$\lambda7136$ & - & $8.9\pm0.8$ & - & $7.9\pm0.6$ & - & - & - & $5.6\pm0.7$ & - \\
{}[\oii]~$\lambda7319$ & - & - & - & - & - & - & - & - & - \\
{}[\oii]~$\lambda7330$ & - & - & - & - & - & - & - & - & - \\
\hline
{\it c}(\hb) ~(mag) & $0.12\pm0.03$ & $0.33\pm0.03$ & $0.7\pm0.1$ & $0.56\pm0.04$ & $0.6\pm0.2$ & $0.1\pm0.2$ & $0.5\pm0.2$ & $0.38\pm0.02$ & $0.18\pm0.01$ \\
$F(\hb)$\tnote{a} & 1.3$\pm$0.1 & $18\pm1$ & 2.6$\pm$0.7 & $23\pm2$ & $4\pm2$ & 1.6$\pm$0.7 & $4\pm2$ & $20\pm1$ & $4.1\pm0.2$ \\
EW$_{\textrm{abs}}$~(\AA) & 0.0 & 0.8 & 0.0 & 0.0 & 0.0 & 0.0 & 0.0 & 0.0 & 1.0 \\
Aperture~(\arcsec) & 10.6 & 19.1 & 6.4 & 12.7 & 12.7 & 6.4 & 12.7 & 31.3 & 6.9 \\
\hline
\end{tabular*}

\begin{tabular}{cccccccccc}
 & & &  &  & &  &  &  &  \\
\end{tabular}

\begin{tabular*}{\textwidth}{@{\extracolsep{\fill}}lccccccccc}
\hline\hline
Line ID & \multicolumn{9}{c}{\hii\ region} \\
        & s2.3 & s2.4 & s2.5 & s3.1 & s3.2 & s3.3 & s3.4 & s3.6 & s3.7 \\
\hline
{}[\oii]~$\lambda3727$ & $262\pm8$ & $330\pm10$ & $260\pm10$ & - & $330\pm10$ & $197\pm6$ & $290\pm10$ & $530\pm40$ & - \\
H$\epsilon$~$\lambda$3970 & $15.9\pm0.6$ & $15\pm1$ & $17\pm4$ & - & $12\pm1$ & - & - & - & - \\
H$\delta$~$\lambda$4102 & $23.6\pm0.8$ & $26\pm1$ & $28\pm4$ & - & $24\pm2$ & $26\pm2$ & - & - & - \\
H$\gamma$~$\lambda$4340 & $45\pm1$ & $47\pm2$ & $47\pm4$ & - & $47\pm2$ & $47\pm2$ & $46\pm6$ & $52\pm8$ & $150\pm30$ \\
{}[\oiii]~$\lambda4363$ & $0.8\pm0.2$ & - & - & - & - & - & - & - & - \\
H$\beta$~$\lambda$4861 & $100\pm3$ & $100\pm3$ & $100\pm4$ & $100\pm10$ & $100\pm3$ & $100\pm3$ & $100\pm4$ & $100\pm7$ & $100\pm20$ \\
{}[\oiii]~$\lambda4959$ & $63\pm2$ & $40\pm1$ & $55\pm2$ & $130\pm20$ & $37\pm1$ & $29\pm1$ & $31\pm2$ & $150\pm9$ & $190\pm30$ \\
{}[\oiii]~$\lambda5007$ & $190\pm5$ & $120\pm3$ & $163\pm6$ & $409\pm40$ & $110\pm3$ & $81\pm2$ & $69\pm3$ & $409\pm20$ & $580\pm80$ \\
{}[\nii]~$\lambda5755$ & $0.52\pm0.07$ & - & - & - & - & - & - & $14\pm2$ & - \\
\hei~$\lambda$5876 & $12.4\pm0.4$ & $10.7\pm0.4$ & $11\pm1$ & - & $9.6\pm0.5$ & $10.1\pm0.5$ & - & - & - \\
{}[\oi]~$\lambda6300$ & $2.8\pm0.1$ & $1.8\pm0.2$ & - & - & - & - & $9\pm1$ & $14\pm2$ & - \\
{}[\siii]~$\lambda6312$ & $1.2\pm0.1$ & $1.4\pm0.2$ & - & - & - & - & - & - & - \\
{}[\oi]~$\lambda6364$ & $1.08\pm0.1$ & $0.9\pm0.2$ & - & - & - & - & - & - & - \\
{}[\nii]~$\lambda6548$ & $14.9\pm0.5$ & $16.4\pm0.5$ & $13\pm1$ & $11\pm4$ & $18.2\pm0.6$ & $17.4\pm0.7$ & $25\pm1$ & $16\pm2$ & - \\
H$\alpha$~$\lambda$6563 & $340\pm10$ & $289\pm9$ & $290\pm10$ & $290\pm30$ & $289\pm9$ & $288\pm9$ & $286\pm10$ & $290\pm20$ & $290\pm40$ \\
{}[\nii]~$\lambda6583$ & $44\pm1$ & $47\pm1$ & $39\pm2$ & $17\pm4$ & $55\pm2$ & $55\pm2$ & $78\pm3$ & $58\pm4$ & $32\pm7$ \\
\hei~$\lambda$6678 & $3.8\pm0.1$ & $3.0\pm0.2$ & $5.1\pm0.8$ & - & $3.4\pm0.3$ & $3.2\pm0.4$ & $4.6\pm1.0$ & - & - \\
{}[\sii]~$\lambda6717$ & $31\pm1$ & $28.1\pm0.8$ & $31\pm2$ & $20\pm5$ & $36\pm1$ & $38\pm1$ & $81\pm3$ & $72\pm5$ & $42\pm8$ \\
{}[\sii]~$\lambda6731$ & $23.6\pm0.8$ & $19.8\pm0.6$ & $21\pm1$ & $22\pm5$ & $26.3\pm0.9$ & $27.1\pm0.9$ & $60\pm2$ & $56\pm4$ & $36\pm8$ \\
\hei~$\lambda$7065 & $2.3\pm0.3$ & $1.5\pm0.2$ & - & - & - & - & - & - & - \\
{}[\ariii]~$\lambda7136$ & $10.4\pm0.4$ & $7.8\pm0.3$ & $9.9\pm0.7$ & - & $7.3\pm0.4$ & $5.4\pm0.5$ & $4\pm1$ & $11\pm2$ & - \\
{}[\oii]~$\lambda7319$ & $2.8\pm0.3$ & $2.4\pm0.2$ & - & - & - & - & - & - & - \\
{}[\oii]~$\lambda7330$ & $2.6\pm0.3$ & $2.1\pm0.2$ & - & - & - & - & - & - & - \\
\hline
{\it c}(\hb) ~(mag) & $0.18\pm0.02$ & $0.51\pm0.01$ & $0.36\pm0.03$ & $0.16\pm0.03$ & $0.25\pm0.02$ & $0.11\pm0.01$ & $0.21\pm0.01$ & $0.4\pm0.02$ & $0.52\pm0.02$ \\
$F(\hb)$\tnote{a}& $146\pm7$ & $111\pm3$ & 9.4$\pm$0.7 & 1.5$\pm$0.2 & $41\pm2$ & $57\pm2$ & 8.2$\pm$0.3 & 3.9$\pm$0.3 & 1.5$\pm$0.2 \\
EW$_{\textrm{abs}}$~(\AA)& 0.0 & 0.9 & 0.0 & 0.0 & 2.2 & 2.8 & 2.2 & 0.0 & 0.0 \\
Aperture~(\arcsec) & 24.0 & 21.0 & 12.7 & 7.3 & 28.1 & 88.5 & 19.0 & 12.7 & 5.3 \\
\hline
\end{tabular*}
\begin{tablenotes}
\item [a] In units of 10$^{-15}$~erg~s$^{-1}$~cm$^{-2}$.
\end{tablenotes}
\end{threeparttable}
\end{table*}

\begin{table*}
\begin{threeparttable}
\caption{\label{line_fluxes2} Continuation of Table~\ref{line_fluxes}.}
\centering
\begin{tabular*}{\textwidth}{@{\extracolsep{\fill}}lccccccccc}
\hline\hline
Line ID & \multicolumn{9}{c}{\hii\ region} \\
        & s3.8 & s4.1 & s4.2 & s4.3 & s4.4 & s5.1 & s5.2 & s5.3 & s5.4 \\
\hline
{}[\oii]~$\lambda3727$ & $310\pm30$ & $197\pm8$ & $330\pm70$ & $310\pm60$ & $259\pm9$ & $540\pm90$ & $229\pm20$ & $280\pm20$ & $300\pm40$ \\
H$\delta$~$\lambda$4102 & - & $25\pm3$ & - & - & $26\pm3$ & - & $28\pm6$ & $30\pm6$ & - \\
H$\gamma$~$\lambda$4340 & $51\pm7$ & $47\pm2$ & - & - & $47\pm2$ & - & $44\pm5$ & $48\pm4$ & $53\pm9$ \\
{}[\oiii]~$\lambda4363$ & - & - & - & - & - & - & - & - & - \\
H$\beta$~$\lambda$4861 & $100\pm6$ & $100\pm3$ & $100\pm10$ & $100\pm10$ & $100\pm3$ & $100\pm10$ & $100\pm5$ & $100\pm4$ & $100\pm9$ \\
{}[\oiii]~$\lambda4959$ & $44\pm4$ & $67\pm2$ & $34\pm8$ & - & $32\pm1$ & $55\pm9$ & $97\pm5$ & $93\pm4$ & $61\pm7$ \\
{}[\oiii]~$\lambda5007$ & $115\pm6$ & $196\pm6$ & $80\pm10$ & $42\pm8$ & $97\pm3$ & $200\pm20$ & $290\pm10$ & $280\pm10$ & $200\pm10$ \\
{}[\nii]~$\lambda5755$ & - & - & - & - & - & - & - & - & - \\
\hei~$\lambda$5876 & - & $12.0\pm0.6$ & - & - & $12.3\pm0.5$ & - & $12\pm1$ & $13\pm1$ & - \\
{}[\oi]~$\lambda6300$ & - & - & - & - & $3.3\pm0.3$ & - & - & - & - \\
{}[\siii]~$\lambda6312$ & - & - & - & - & - & - & - & - & - \\
{}[\oi]~$\lambda6364$ & - & - & - & - & $1.2\pm0.3$ & - & - & - & - \\
{}[\nii]~$\lambda6548$ & $6\pm2$ & $11.1\pm0.5$ & $20\pm4$ & $25\pm5$ & $16.7\pm0.6$ & $11\pm3$ & $8\pm1$ & $10.0\pm1.0$ & $15\pm2$ \\
H$\alpha$~$\lambda$6563 & $290\pm30$ & $288\pm10$ & $290\pm50$ & $280\pm50$ & $289\pm9$ & $290\pm50$ & $290\pm20$ & $290\pm20$ & $290\pm30$ \\
{}[\nii]~$\lambda6583$ & $23\pm3$ & $30\pm1$ & $55\pm10$ & $60\pm10$ & $47\pm2$ & $42\pm7$ & $23\pm2$ & $26\pm2$ & $46\pm4$ \\
\hei~$\lambda$6678 & - & $3.7\pm0.3$ & - & - & $3.6\pm0.2$ & - & - & $4.3\pm0.8$ & - \\
{}[\sii]~$\lambda6717$ & $30\pm4$ & $19.3\pm0.7$ & $49\pm9$ & $70\pm10$ & $37\pm1$ & $60\pm10$ & $17\pm2$ & $26\pm2$ & $47\pm4$ \\
{}[\sii]~$\lambda6731$ & $22\pm3$ & $13.2\pm0.5$ & $33\pm6$ & $51\pm9$ & $25.4\pm0.9$ & $37\pm7$ & $14\pm2$ & $18\pm1$ & $31\pm3$ \\
\hei~$\lambda$7065 & - & $2.1\pm0.4$ & - & - & $2.9\pm0.4$ & - & - & - & - \\
{}[\ariii]~$\lambda7136$ & - & $8.2\pm0.5$ & $18\pm4$ & - & $6.0\pm0.4$ & $11\pm3$ & $11\pm1$ & $10\pm1$ & $9\pm1$ \\
{}[\oii]~$\lambda7319$ & - & - & - & - & $2.2\pm0.3$ & - & - & - & - \\
{}[\oii]~$\lambda7330$ & - & - & - & - & $2.1\pm0.3$ & - & - & - & - \\
\hline
{\it c}(\hb) ~(mag) & $0.06\pm0.02$ & $0.19\pm0.02$ & $0.4\pm0.2$ & $0.3\pm0.2$ & $0.13\pm0.02$ & $0.4\pm0.2$ & $0.28\pm0.1$ & $0.34\pm0.08$ & $0.29\pm0.08$ \\
$F(\hb)$\tnote{a} & 2.1$\pm$0.1 & $33\pm2$ & $6\pm3$ & $5\pm2$ & $52\pm3$ & $7\pm3$ & $11\pm3$ & $18\pm3$ & $17\pm3$ \\
EW$_{\textrm{abs}}$~(\AA)  & 0.0 & 1.0 & 1.0 & 1.0 & 1.3 & 0.0 & 0.0 & 0.0 & 0.0 \\
Aperture~(\arcsec) & 10.0 & 15.9 & 11.1 & 9.5 & 21.2 & 21.2 & 7.4 & 14.3 & 47.7 \\
\hline
\end{tabular*}
\begin{tablenotes}
\item [a] In units of 10$^{-15}$~erg~s$^{-1}$~cm$^{-2}$.
\end{tablenotes}
\end{threeparttable}
\end{table*}

\section{\texorpdfstring{\hii\ region catalogues for NGC~628}{H II region catalogues for NGC~628}}
\label{sec:appenN628}
There are many published catalogues of the \hii\ regions of the spiral  NGC~628. This is the main reason for including  this galaxy in our pilot study. 
We have made a selection of published catalogues, including those of  \citet{Fathi2007}, \citet{Rosales-Ortega11}, \citet{Cedres2012}, \citet{RN18}, \citet{Santoro2022}, \citet{Groves2023} and \citet{Congiu2023}. 
There are a few more available in the literature \citep[e.g.][]{vonHippel, KennicuttHodge80}, but this selection is sufficient to illustrate how the cataloguing methodology (last column in Table~\ref{tab:catalogues_stat}) has a strong impact on the number of catalogued regions and their directly measured properties (such as \ha\ luminosity and projected size), as well as on derived properties such as the rms or average electron density, which is the focus of this work.

\begin{figure*}
\centering  
\includegraphics[width=0.66\linewidth]{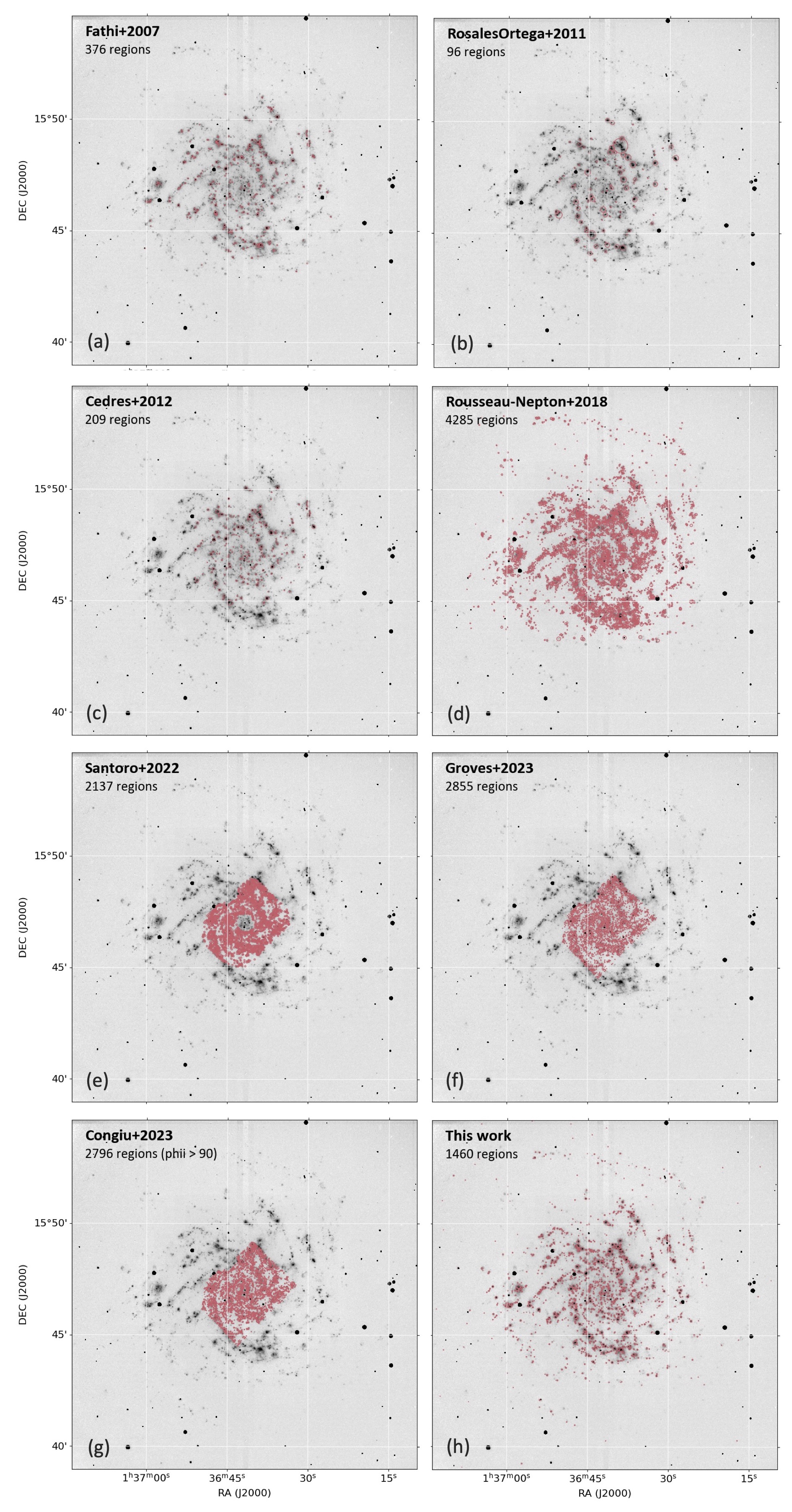}
\caption{Greyscale continuum-subtracted \ha\ image of NGC~628. In each panel, the red circles mark the location of the \hii\ regions included in the catalogue derived by (a) \citet{Fathi2007}, (b) \citet{Rosales-Ortega11}, (c) \citet{Cedres2012}, (d) \citet{RN18}, (e) \citet{Santoro2022}, (f) \citet{Groves2023}, (g) \citet{Congiu2023}, and (h) by ourselves in this work, as indicated on the top-left corner of each panel, together with the total  number of catalogued \hii\ regions. The red circles are scaled to the effective radius of the \hii\ region as derived from the corresponding catalogue, except for  \citet{Santoro2022}, who do not provide region sizes.}
\label{fig:finders_all}
\end{figure*}

Table~\ref{tab:catalogues_stat} clearly shows substantial differences among the various \hii\ region catalogues, starting with the total number of identified regions. The three catalogues based on MUSE observations\footnote{In this work, we restrict the \citet{Congiu2023} sample to objects with {\tt p\_hii > 90}, corresponding to sources with the highest probability of being genuine \hii\ regions, as defined by the authors.} \citep{Santoro2022, Groves2023, Congiu2023} contain more than 2000 \hii\ regions each, whereas the earlier works by \citet{Fathi2007}, \citet{Rosales-Ortega11}, and \citet{Cedres2012} contain fewer than 400 regions, even though the former were obtained from observations covering only the inner part of the disc, as can be seen in Fig.~\ref{fig:finders_all}.

\begin{figure*}
    \centering
    \includegraphics[width=0.65\linewidth]{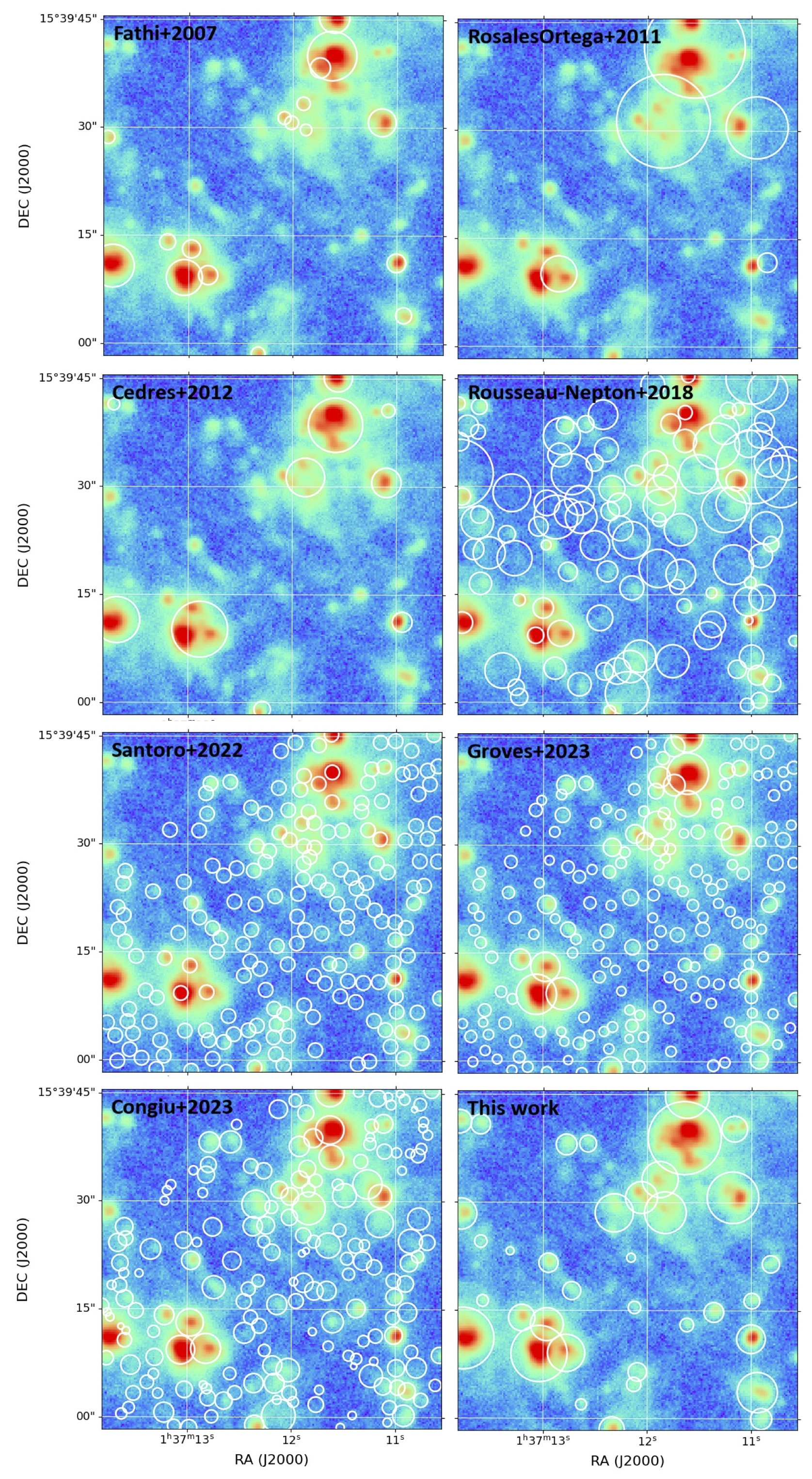}
    \caption{Section of the continuum-subtracted \ha\ image of NGC~628, displaying
     the catalogued \hii\ regions (white circles), in order to better appreciate the differences between the catalogues. The circles are scaled to the effective radius of the \hii\ regions, as derived from the corresponding catalogue, except for \citet{Santoro2022}, who do not provide region sizes.}
    \label{fig:finders_zoom_all}
\end{figure*}

Fig.~\ref{fig:finders_zoom_all} presents a zoom-in of the continuum-subtracted \ha\ image of the galaxy, and offers qualitative explanations for part of these differences in the total number of regions. On the one hand, bright \ha\ emission knots are often decomposed into multiple individual \hii\ regions in the catalogues of \citet{RN18}, \citet{Santoro2022}, \citet{Groves2023}, and \citet{Congiu2023} while other authors, including ourselves, tend to treat them as single regions (see e.g. knots in the top right of each panel). On the other hand, Fig.~\ref{fig:finders_zoom_all} also indicates that the total number may also differ due to: (1) the non-detection of low-luminosity \hii\ regions in the catalogues of \citet{Fathi2007} and \citet{Cedres2012}, presumably due to the lower signal-to-noise ratio of their observational data, collected with smaller aperture, 1–2.5~m-class telescopes, or (2) the exclusion of zones with low surface brightness. This is the case in our catalogue, in which low surface brightness regions are considered part of the diffuse ionized gas  (see Sect.~\ref{sec:catalogue}), while some of these appear to be included as \hii\ regions by \citet{RN18}, \citet{Santoro2022}, \citet{Groves2023}, and \citet{Congiu2023} in their catalogues.
These qualitative differences may also account for the lower median \ha\ luminosities reported in the catalogues by \citet{Santoro2022}, \citet{Groves2023}, and \citet{Congiu2023}, which are at least 0.2–0.4~dex lower than derived from our catalogue (noting that our values are uncorrected for extinction). The same applies to the median equivalent radius in the catalogues of \citet{Groves2023} and \citet{Congiu2023}.

The comparison with the catalogue of \citet{RN18} is less straightforward. While their reported median \ha\ luminosity and equivalent radius are similar to ours, a detailed inspection of their panel in Fig.~\ref{fig:finders_zoom_all} reveals clear  differences in the determination of \hii\ region sizes compared to other studies. Their methodology is based on what the authors term the  “zone of influence” for each detected peak. 
In the zoomed-in area shown in Fig.~\ref{fig:finders_zoom_all}, this approach appears to identify a significant  number of larger, low surface brightness \hii\ regions, presumably also  of correspondingly low \ha\ luminosity, than those catalogued by other authors for the same area.

\begin{figure*}
    \centering
    \includegraphics[width=\linewidth]{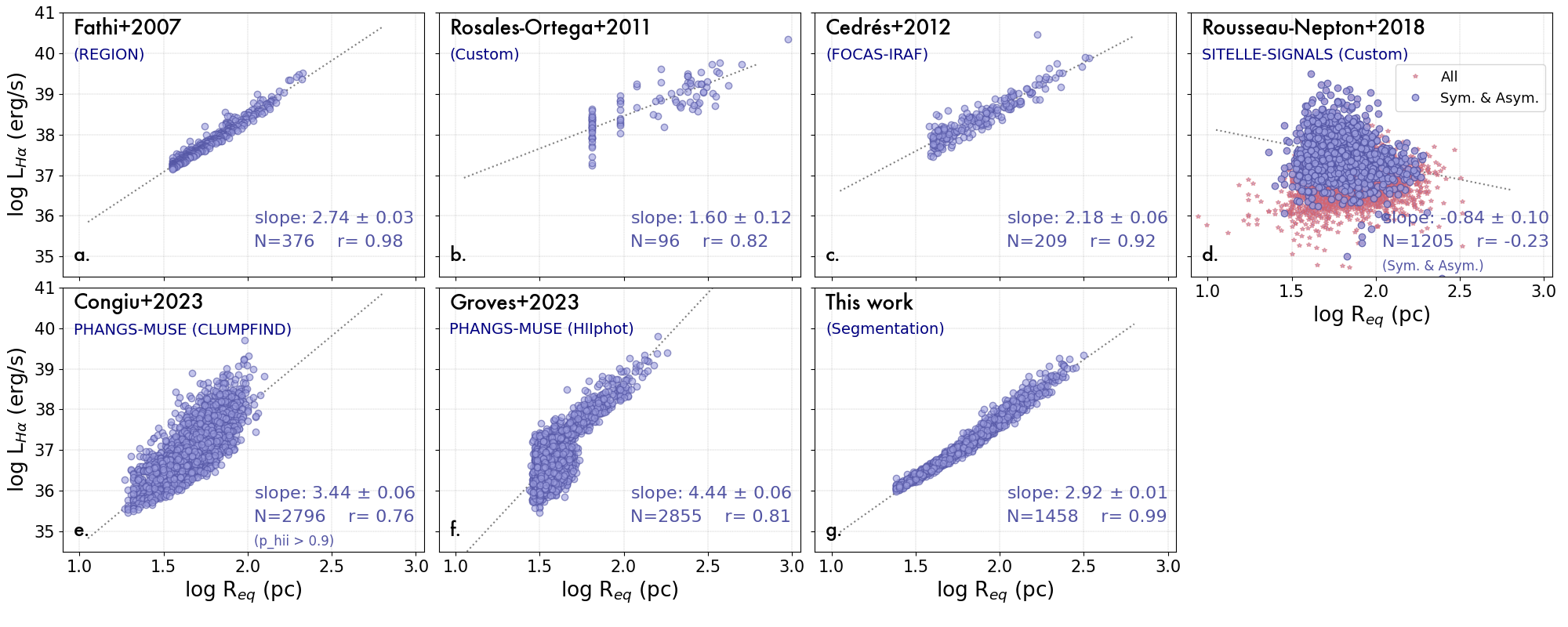}
    \caption{Logarithm of the \ha\ luminosity (in erg~s\me) as a function of the logarithm of the equivalent radius (in parsecs) for the \hii\ regions of the NGC~628 catalogues published by  (a) \citet{Fathi2007}, (b) \citet{Rosales-Ortega11}, (c) \citet{Cedres2012}, (d) \citet{RN18}, (e) \citet{Congiu2023}, (f) \citet{Groves2023}, and (g) by ourselves in this work. The dotted  lines show the best linear fit to the data points in each catalogue, with the derived slope, number of data points and correlation coefficient in the bottom-right corner of the corresponding panel. For the catalogues derived by \citet{RN18} and \citet{Congiu2023}, only regions classified as symmetrical or asymmetrical, and having {\tt p\_hii}$> 90$ are considered for the linear fit, respectively. }
    \label{fig:logL-logR_all}
\end{figure*}

Methodological differences in the cataloguing process generally have a much stronger impact on the measured radii of \hii\ regions than on their luminosities, since the central part of the region contributes the most to the total $L_{\mathrm{H}\alpha}$. 
As a result, the slopes of the luminosity functions tend to be relatively robust against such differences \citep[e.g.][]{Santoro2022}. Figure~\ref{fig:logL-logR_all} shows the 
$\log~L_{\mathrm{H}\alpha} - \log~R_{\mathrm{eq}}$ relation for the compiled catalogues (excluding that of \citealt{Santoro2022}, which does not report radii). 
The relation varies significantly from one catalogue to another, although in all cases, with the exception of the relation based on the \citet{RN18} catalogue, the expected positive correlation between these two parameters is found (Eq.~\ref{LVec}).
A single linear fit to the relations reveals slopes that differ widely, ranging from 1.60 \citep{Rosales-Ortega11} to 4.44 \citep{Groves2023}, as does the scatter. In the case where neither \ne\ nor $\phi$ depend on $R_{\mathrm{eq}}$, the slope is expected to be 3. Only the catalogues derived in \citet{Fathi2007} and in this work yield slopes close to that value, with $2.74\pm0.03$ and $2.92\pm0.01$, respectively.

We remark that the $\log~L_{\mathrm{H}\alpha} - \log~R_{\mathrm{eq}}$ relations derived by \citet{Groves2023} and \citet{Congiu2023} differ significantly, despite the fact that both catalogues have been produced by the same group using the same MUSE data. 
In particular, regions with $\log~L_{\mathrm{H}\alpha} \lesssim 37.4$
show a completely different distribution in the $\log~L_{\mathrm{H}\alpha} - \log~R_{\mathrm{eq}}$ plane. 
The peculiar relation observed for the catalogue of \citet{RN18} in Fig.~\ref{fig:logL-logR_all} likely reflects the methodological differences in the estimation of \hii\ region sizes discussed earlier in this section: in their catalogue the largest regions are not necessarily the most luminous ones, even when the analysis is restricted to those classified by the authors as symmetric or asymmetric.

\begin{figure*}
    \centering
    \includegraphics[width=\linewidth]{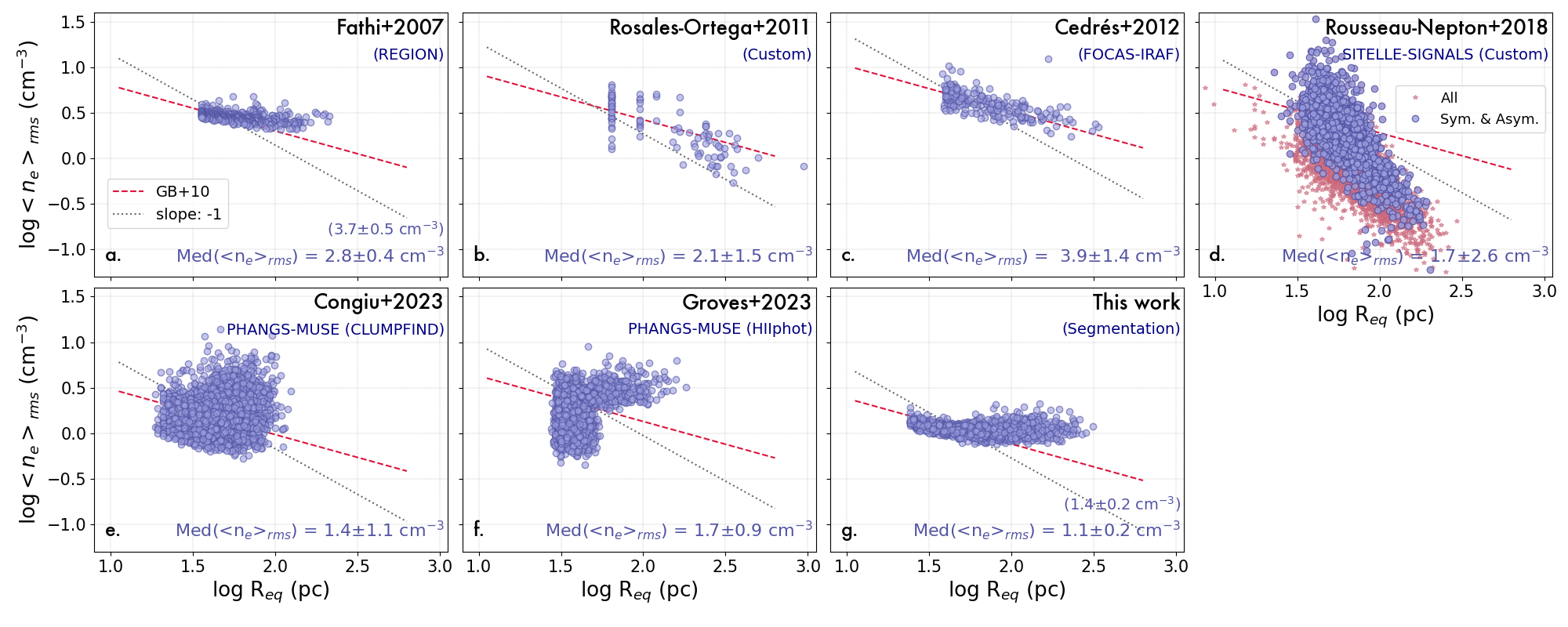}
    \caption{Logarithm of the rms electron density (in cm$^{-3}$) as a function the logarithm of the equivalent radius (in parsecs) for the \hii\ regions of the NGC~628 catalogues published by  (a) \citet{Fathi2007}, (b) \citet{Rosales-Ortega11}, (c) \citet{Cedres2012}, (d) \citet{RN18}, (e) \citet{Congiu2023}, (f) \citet{Groves2023},  and (g) by ourselves in this work. The red dashed and black dotted straight lines show the \nerms$\propto R^{-0.5}$ and \nerms$\propto R^{-1}$ relations derived by \citet{GB10} and \citetalias{HH09}, respectively. The lines have been shifted vertically in each panel to match approximately the location of the data points with $\log R_{\mathrm{eq}}\simeq1.7$. The median \nerms\ value is shown at the bottom of each panel. Values in parentheses indicate the median after applying a uniform extinction correction to the \ha\ luminosities, based on the median {\it c}(\hb) of the \hii\ regions in NGC~628, for catalogues that do not include  extinction correction.}
    \label{fig:logN-logR_all}
\end{figure*}

Fig.~\ref{fig:logN-logR_all} shows the rms electron density,  \nerms, as a function of $\log~R_{\mathrm{eq}}$. The \nerms\ values one derives depend directly on $L_{\mathrm{H}\alpha}$ and $R_{\mathrm{eq}}$ (see Eq.~\ref{eq:ne_rms}). Therefore, the relationships $\log~\nerms - \log~R_{\mathrm{eq}}$ shown in Fig.~\ref{fig:logN-logR_all} reflect the differences in the $\log~L_{\mathrm{H}\alpha} - \log~R_{\mathrm{eq}}$ relation already analysed in Fig.~\ref{fig:logL-logR_all}. We have included in the figure a dashed and dotted straight line  showing the expected slopes of the \nerms$\propto R^{-0.5}$ and \nerms$\propto R^{-1}$ relations derived by \citet{GB10} and \citetalias{HH09}, respectively (shifted vertically in each panel to match approximately the location of the data points with $\log R_{\mathrm{eq}}\simeq1.7$). Although a negative relation is observed with most of the catalogues (i.e.~smaller \hii\ regions tend to have higher average electron densities, except in \citealt{Groves2023}), the slope and average absolute value of \nerms\ clearly  depend on the methodology used to derive the luminosities and projected sizes of the \hii\ regions. The largest median value is found in the \citet{Cedres2012} catalogue, with $3.8$~cm$^{-3}$, whereas the smallest corresponds to \citet{Congiu2023}, with $1.3 \pm 1.0$~cm$^{-3}$, similar to our median value of $1.4$~cm$^{-3}$ after applying a uniform correction based on the average extinction of the \hii\ regions in NGC~628, although with a considerably smaller dispersion (0.2 compared to 1.0).


\bsp	
\label{lastpage}
\end{document}